\documentclass[aps,prc,twocolumn,groupedaddress]{revtex4}

\usepackage{amsmath,amssymb}
\usepackage[dvipdfm]{graphicx,color}
\usepackage{subfigure}
\usepackage{longtable}

\begin{document}


\title{Dinucleon correlation of $^9$Li, $^{10}$Be, and $^{9,10}$C}


\author{Fumiharu Kobayashi and Yoshiko Kanada-En'yo}
\affiliation{Department of Physics, Kyoto University, Kyoto 606-8502, Japan}


\date{\today}

\begin{abstract}
We study the dinucleon (dineutron and diproton) correlation 
of the ground states of $^9$Li, $^{10}$Be, and $^{9,10}$C. 
We assume an $\alpha+t$ core for $^9$Li, 
an $\alpha+\alpha$ core for $^{10}$Be and $^{10}$C, 
and an $\alpha+^3$He core for $^9$C, 
and investigate the effect of core structure changes on 
the degree of dineutron formation and spatial expansion from the core. 
For $^9$Li, $t$ cluster breaking in the core significantly enhances 
the dineutron component inside the nuclei. 
Moreover, its component markedly depends on the strength of the spin-orbit interaction 
since a dineutron is fragile and dissociates readily due to the spin-orbit interaction. 
Compared with $^9$Li, the dineutron of $^{10}$Be dissociates largely 
due to the stronger spin-orbit attraction from the $\alpha+\alpha$ core than the $\alpha+t$ core. 
We also investigate diproton features in $^{9,10}$C, the mirror nuclei of $^9$Li and $^{10}$Be, respectively,
and compare them with the dineutron features of $^9$Li and $^{10}$Be. 
No qualitative difference is observed between diprotons and dineutrons in the degree of formation at the surface, 
but a quantitative difference may exist in the degree of dinucleon-size change 
far from the core. 
\end{abstract}


\maketitle

\section{Introduction}
\label{sec:introduction}

Recently, neutron-rich nuclei have been investigated in detail 
and they have been found to play a major role in many exotic phenomena. 
Dineutron correlation is one of the most interesting topics in the physics of neutron-rich nuclei. 
A dineutron is a spatially compact spin-singlet pair of two neutrons. 
Although two neutrons are not bound in free space, 
it is theoretically suggested that dineutron correlation may be enhanced 
and a compact dineutron is formed in some situations 
such as in a low-density region of nuclear matter \cite{baldo90, matsuo06} 
and in the neutron-halo and neutron-skin regions of finite neutron-rich nuclei 
\cite{bertsch91, zhukov93, arai01, descouvemont03, matsuo05, hagino05, descouvemont06, hagino07}. 
Since a spin-singlet pair of two neutrons is unbound, 
a dineutron is fragile and its features, 
such as the degree of formation and size, can readily change depending on the circumstance. 
For example, in neutron matter, 
it was suggested that the strength of the spatial correlation between two neutrons 
depends markedly on the neutron density. 
The transition between the spatially weak pairing (BCS-like (Bardeen-Cooper-Schrieffer-like) pairing) 
and the spatially strong pairing (boson-like pairing) under a changing neutron density 
was discussed in relation to the BCS-BEC (Bose-Einstein condensate) crossover \cite{matsuo06}. 
In finite nuclei, especially in a neutron-rich nucleus such as $^{11}$Li, 
dineutron correlation is markedly enhanced at the nuclear surface
and the relative distance between two neutrons in a dineutron becomes typically $2-3$ fm 
such that a dineutron may be regarded as a kind of cluster (boson) since the distance of separation is much smaller than the size of the total system. 
As is well known, finite nuclei show various structures of deformation, clusterization, neutron-halo or neutron-skin, etc. 
Such diverse structures may affect dineutron correlation. 
To clarify the universal properties of dineutron correlation in finite nuclei, 
we constructed a model of a dineutron condensate (DC) wave function \cite{kobayashi11}, 
which describes the behavior of a dineutron around a core in detail. 
This model requires no assumptions about the core structure
so that this model is useful to systematically investigate dineutron features in typical neutron-rich nuclei. 
We applied this model to $^{10}$Be ($2\alpha$ core$+2n$)
and investigated dineutron formation around the $2\alpha$ core \cite{kobayashi11}. 
We found that a compact dineutron is formed just at the nuclear surface 
and it spreads somewhat far from the core keeping a compact size in the $^{10}$Be ground state. 
We suggested that a dineutron would be formed in typical nuclei. 

In this work, we investigate dineutron correlation in the $^9$Li ground state ($^7$Li core$+2n$) 
and clarify the effect of $^7$Li core structure change on dineutron correlation. 
We mainly discuss dineutron features and, 
in particular, the degrees of formation and spatial expansion from the core. 
Free $^7$Li is well described by the $\alpha+t$ cluster structure. 
However, the $t$ cluster in the $^7$Li core of $^9$Li is deeply bound and more or less fragile, 
so it is expected that $t$ cluster breaking would be an important aspect of the $^7$Li core description. 
This is rather different from the $\alpha+\alpha$ core in $^{10}$Be 
where $\alpha$ cluster breaking is minor. 
We describe the $^7$Li core as the $\alpha+t$ structure 
and take into account $t$ cluster breaking in addition to its spatial development
as a change in the $^7$Li core structure.  
We also investigate the effect of such core structure changes on dineutron features around the core 
by applying a DC wave function. 
We also compare the dineutron behavior of $^9$Li and $^{10}$Be 
to clarify the difference between dineutron features 
around the $\alpha+t$ and $\alpha+\alpha$ cores. 

Furthermore, we investigate the spatial correlation between two protons in the spin-singlet channel of proton-rich nuclei, 
a phenomenon that is called diproton correlation.
Since the Coulomb repulsive force acts between two protons, 
diproton correlation may vary qualitatively or quantitatively from dineutron correlation. 
Experimental and theoretical studies suggest that 
diproton correlation is just as likely as dineutron correlation
in spite of the Coulomb interaction between two protons
\cite{goldansky60, grigorenko00, grigorenko01, giovinazzo02, grigorenko03, garrido04, 
grigorenko07, oishi10}. 
In this work, we also investigate the difference between dineutron and diproton features 
by comparing the diproton correlation of $^{9,10}$C 
with the dineutron correlation of $^9$Li and $^{10}$Be, since $^{9,10}$C are the mirror nuclei of $^9$Li and $^{10}$Be, respectively. 

The content of this paper is as follows. 
In Sec.~\ref{sec:framework}, 	
we describe the framework used in the present study. 
In Sec.~\ref{sec:dineutron}, 
we examine dineutron correlation results in the ground states of $^9$Li and $^{10}$Be. 
We first evaluate the dineutron probability for $^9$Li 
and consider the effect of core structure change on dineutron features. 
Next, we compare the dineutron features of $^9$Li with those of $^{10}$Be
and demonstrate that their differences arise from the core structure.
In Sec.~\ref{sec:diproton}, 
we discuss the diproton features of $^9$C and $^{10}$C
and compare them with dineutron features in their mirror nuclei $^9$Li and $^{10}$Be, respectively. 
Finally, we provide a summary of the present work in Sec.~\ref{sec:summary}.

\section{Framework}
\label{sec:framework}

In this section, we first describe the framework used to investigate the dineutron correlation of $^9$Li. 
The framework used for both $^{10}$Be and $^{9,10}$C is similar to that employed for $^9$Li, 
and only a brief explanation is subsequently added concerning these latter nuclei. 

To investigate dineutron correlation in the $^9$Li ground state, 
we superpose two kinds of wave functions: one to describe the main structure of the ground state 
and the other to efficiently describe the dineutron correlation about the $^7$Li core. 
For the former, we use the $^6$He$+t$ cluster wave function as 
proposed in Ref.~\cite{enyo12}, 
which was previously used to describe 
the $^9$Li system as $^6$He ($\alpha+2n[(0p)^2]$) plus $t$ clusters. 
By using such a wave function, 
the low-lying energy spectrum of $^9$Li can be well reproduced. 
For the latter, we use a DC wave function having an $\alpha+t$ core. 
This DC wave function can describe the detailed features of a dineutron about the core, 
that is, the size change of the dineutron and 
the expansion of the dineutron distribution from the core \cite{kobayashi11, kobayashi12, kobayashi13}. 
By superposing these wave functions, 
we investigate the competition between dineutron formation 
(wherein two neutrons couple to a spin singlet with spatial correlation)
and its dissociation at the surface due to spin-orbit interaction
(wherein two neutrons are in the major shell independently) 
and the expansion of the dineutron distribution far from the core. 
The $^6$He$+t$ cluster wave function 
can describe spin-singlet pair formation 
and pair dissociation to the $(0p_{3/2})^2$ configuration of two neutrons 
just at the surface of the $\alpha+t$ core. 
Using the DC wave function, 
we consider the dineutron spatial expansion from the core and the dineutron-size change. 
We investigate the effect of the $\alpha+t$ core structure change on
such dineutron features as the dineutron formation and dissociation near the surface 
and its expansion from the core. 
For this purpose, we take into account $t$ cluster breaking and development in the $\alpha+t$ core 
of the DC wave function as well as that in the $^6$He$+t$ cluster wave function. 
For $t$ cluster breaking, we introduce the method 
proposed by Itagaki {\it et al} to break a cluster to gain the spin-orbit energy \cite{itagaki05}. 

In the following, 
we first describe the details of the $^6$He$+t$ cluster wave function 
and the $^9$Li DC wave function without $t$ cluster breaking.
Subsequently, we explain $t$ cluster breaking in the core.

\subsection{The $^6$He+$t$ cluster wave function}
\label{sec:extended_cluster_wf}

\begin{figure}[b]
\includegraphics[scale=0.3]{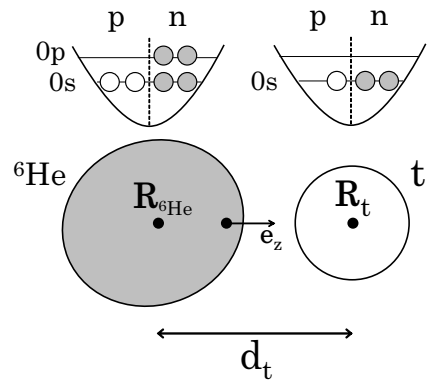}
\caption{A schematic representation of the $^6$He$+t$ cluster wave function. 
$^6$He is described with $\pi[(0s)^2]\nu[(0s)^2 (0p)^2]$ configurations around $\boldsymbol{R}_{^6{\rm He}}$. 
The $t$ cluster contains a spin-up proton and spin-up and spin-down neutrons in the $0s$ orbit
around $\boldsymbol{R}_t$. }
\label{fig:He6+t_cluster_wf}
\end{figure}

First, we explain the simple $^6$He$+t$ cluster wave function 
proposed in Ref.~\cite{enyo12} without $t$ cluster breaking. 
For the $^6$He+$t$ cluster wave function, 
the $^9$Li system consists of the $^6$He cluster and the $t$ cluster, 
with the cluster wave functions being described by the Bloch-Brink wave function \cite{brink}. 
The $^6$He cluster is composed of a $(0s)^4$ $\alpha$ cluster and two valence neutrons
in harmonic oscillator (H.O.) $(0p)^2$ configurations around the $\alpha$. 
The $^6$He$+t$ cluster wave function, $\Phi_{^6{\rm He}+t}$, is as follows. 
\begin{align}
\Phi_{^6{\rm He}+t} (\kappa, d_t)
= \mathcal{A} \big\{ &
\Phi_{^6{\rm He}}(\kappa, \boldsymbol{R}_{^6{\rm He}} = -d_t / 3 \ \boldsymbol{e}_z)  \nonumber \\
& \times \Phi_{t}(\boldsymbol{R}_t = 2 d_t/3 \ \boldsymbol{e}_z) \big\}, 
\label{eq:He6+t_cluster_wf}
\end{align}
where $\Phi_{^6{\rm He}, t}$ are the wave functions of the $^6$He and $t$ clusters. 
The $t$ cluster is composed of a spin-up proton and spin-up and spin-down neutrons. 
Here, the $t$ cluster is an ideal $(0s)^3$ $t$ cluster, 
which we define as a $t$ cluster. 
The spatial component of the single-particle wave function, $\phi$, is described with the Gaussian wave packet as 
\begin{equation}
\phi (\boldsymbol{r}; \nu, \boldsymbol{R}) 
\propto \exp \big[ - \nu (\boldsymbol{r}- \boldsymbol{R})^2 \big]. 
\label{eq:sp_wf}
\end{equation}
In the $^6$He cluster, two valence neutrons are in the $0p$ configurations 
described by a shifted Gaussian from the $\alpha$, 
and $\kappa$ assigns the $0p$ configuration of two valence neutrons around the $\alpha$. 
$\boldsymbol{R}_{^6{\rm He}, t}$ are the center of mass of the Gaussian center parameters 
in the $^6$He and $t$ clusters, 
and the $^6$He and $t$ clusters are separated by a distance $d_t$ in the $z$-direction. 
The Gaussian width parameter, $\nu$, is chosen as 
$\nu = 0.235$ fm$^{-2}$ for all nucleons. 
A schematic representation of the $^6$He$+t$ cluster wave function is shown 
in Fig.~\ref{fig:He6+t_cluster_wf}.
By using the $^6$He$+t$ cluster wave function, 
we describe the main structure of the $^9$Li ground state. 
This wave function can describe two valence neutrons in H.O. $(0p)^2$ configurations, 
and, therefore, the mixing of configuration $\kappa$ 
describes the competition between dineutron formation  
(wherein two neutrons couple to the spin singlet in the $0p$ shell) 
and dineutron dissociation 
(wherein two neutrons occupy the $(0p_{3/2})^2$ to gain the spin-orbit interaction) 
just at the surface.

\subsection{The $^9$Li DC wave function}
\label{sec:DC_wf}

\begin{figure}[b]
\includegraphics[scale=0.35]{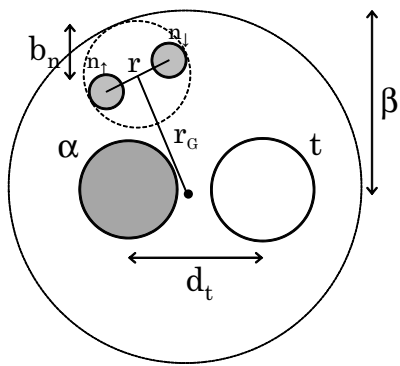}
\caption{A schematic representation of the $^9$Li DC wave function. 
$b_n$ stands for the dineutron size 
and $\beta$ stands for the dineutron expansion from the core. 
The core is described with the $\alpha$ and $t$ clusters separated by $d_t$. }
\label{fig:DC_wf}
\end{figure}

In addition to the $^6$He$+t$ cluster wave functions, 
we superpose the DC wave functions \cite{kobayashi11} for $^9$Li 
to describe dineutron correlation around the $\alpha+t$ core. 
The present DC wave function describes the system composed of the $\alpha+t$ core and one dineutron
of a pair of spin-up and spin-down neutrons. 
For the $^9$Li DC wave function, the dineutron is distributed around the core spherically 
changing its size and spatial expansion from the core as follows. 
\begin{align}
\Phi_{\rm DC}(d_t, \beta, b_n) = &\ \mathcal{A} \{ \Phi_{\alpha+t}(d_t) 
\times \Phi_{2n_{\rm DC}} (\beta, b_n) \}
\label{eq:DC_wf} \\ 
\Phi_{2n_{\rm DC}} (\beta, b_n) = &\ \mathcal{A} \{ \psi_r (b_n) \psi_G(\beta) \ 
\chi_{\uparrow}(n1) \chi_{\downarrow}(n2) \}, 
\label{eq:DC_wf_dineutron} \\
& \psi_r (\boldsymbol{r};b_n) \propto
\exp \left[ - \frac{r^2}{4b_n^2} \right], 
\label{eq:DC_wf_rel} \\
& \psi_G (\boldsymbol{r}_G;\beta) \propto 
\exp \left[ - \frac{r_G^2}{\beta^2} \right]. 
\label{eq:DC_wf_com}
\end{align}
$\Phi_{\alpha+t, 2n_{\rm DC}}$ are the wave functions of the $\alpha+t$ core and the dineutron. 
The dineutron wave function, $\Phi_{2n_{\rm DC}}$, 
is composed of the relative and center-of-mass wave functions of the two neutrons forming the dineutron, 
$\psi_r$ and $\psi_G$, where 
$\chi_{\uparrow/\downarrow}$ are the spin-up and spin-down wave functions.
The coordinate $\boldsymbol{r} = \boldsymbol{r}_{n1} - \boldsymbol{r}_{n2}$ 
is a distance relative to the two neutrons in the dineutron, given by the coordinates $\boldsymbol{r}_{n1,n2}$ , 
while $\boldsymbol{r}_G = (\boldsymbol{r}_{n1} + \boldsymbol{r}_{n2})/2$ 
is their center-of-mass relative to the origin. 
For the DC wave function, 
a spin-singlet dineutron is composed of two neutrons in the relative $s$ wave 
and it is distributed about the origin in the $S$ wave. 
In the relative wave function, $\psi_r$, 
two neutrons have the same size parameter, $b_n$ ($\nu = (2b_n^2)^{-1}$ in Eq.~(\ref{eq:sp_wf})), 
characterizing the relative distance between two neutrons, 
that is, the dineutron size.  
In the center-of-mass wave function, $\psi_G$, 
the spatial expansion of the dineutron is characterized by the parameter $\beta$. 
(By definition, $\beta > b_n$ as shown in Ref.~\cite{kobayashi11}.)
A schematic representation of the $^9$Li DC wave function is shown in Fig.~\ref{fig:DC_wf}. 

In the DC wave function, 
we change these characteristic parameters $(\beta, b_n)$ to various values 
to evaluate detailed dineutron behavior. 
When $b_n$ is small, the distance between two neutrons is small 
and a compact dineutron is formed. 
On the other hand, when $b_n \approx \beta$, 
the DC wave function corresponds to the state 
where each valence neutron is in the $s$ wave with the width $b_n \approx \beta$
around the origin, as shown in Ref.~\cite{kobayashi11}. 
Two neutrons in the $s$ wave no longer exhibit spatial correlation 
and the dineutron disperses. 
Therefore, 
the DC wave function can describe dineutron formation and dispersion by changing the size parameter $b_n$. 
The parameter $\beta$ corresponds to the degree of dineutron expansion from the core. 
When $\beta$ is small, the dineutron is distributed near the core, 
and, when $\beta$ is large, the dineutron distribution is widely expanded from the core. 
In such a way, the parameters $b_n$ and $\beta$ characterize the dineutron features of 
size and distribution about the core. 
We superpose DC wave functions consisting of various sets of $(\beta, b_n)$ 
to describe a system containing a dineutron having various sizes and spatial distributions about the core. 

For the $^9$Li DC wave function, 
the core is described by the $\alpha$+$t$ cluster wave function, $\Phi_{\alpha+t}$. 
The $\alpha$ and $t$ clusters, separated by a distance $d_t$, are described with the ideal $(0s)^4$ and $(0s)^3$ configurations, respectively.
The center of mass of the $^7$Li core is located at the origin
and we omit the core recoil with respect to dineutron motion in the DC wave function. 
The Gaussian width of nucleons in the $\alpha+t$ core is fixed to $\nu = 0.235$ fm$^{-2}$.

\subsection{Development and breaking of the $t$ cluster}
\label{sec:t_breaking}

Here, we examine the $\alpha+t$ core structure change. 
In the $^6$He$+t$ cluster wave function and the $^9$Li DC wave function explained above, 
the system contains the $\alpha+t$ core, 
and the $\alpha$-$t$ relative distance is characterized by $d_t$. 
To describe the $^9$Li ground state, 
we superpose wave functions with various $d_t$ values 
and take into account the spatial development of the $t$ cluster in the $\alpha+t$ core. 

We also consider $t$ cluster breaking in the $\alpha+t$ core as the core structure change. 
While the $\alpha+t$ threshold energy in free $^7$Li is $2.47$ MeV, 
the $^6$He$+t$ threshold energy in $^9$Li is $7.59$ MeV. 
The $t$ cluster is deeply bound in $^9$Li 
and $t$ cluster breaking is expected at the surface,
especially because spin-orbit interaction to occupy the $0p_{3/2}$ orbits 
is important in $^9$Li. 
Thus, we introduce $t$ cluster breaking in the core in addition to $t$ cluster development, 
and we also discuss the effect of core cluster breaking on dineutron features. 
Here, we examine the method by which $t$ cluster breaking is accounted for. 

In a Gaussian-type, single-particle wave function [Eq.~(\ref{eq:sp_wf})], 
the real and imaginary parts of the center parameter, $\boldsymbol{R}$, correspond to 
the mean position and momentum of the particle, respectively.
\begin{align}
\langle \boldsymbol{r} \rangle = &\ {\rm Re} [\boldsymbol{R}], \label{eq:coordinate_sp} \\
\langle \boldsymbol{p} \rangle = &\ 2\hbar \nu {\rm Im}[\boldsymbol{R}]. \label{eq:momentum_sp}
\end{align}
For the $^6$He$+t$ cluster 
and $^9$Li DC wave functions described in the previous sections, 
the $t$ cluster is described using the ideal $(0s)^3$ configuration 
and the nucleons located there have the same real center parameter. 
We extend this framework to address breaking of the ideal $(0s)^3$ $t$ cluster 
by giving the imaginary part (i.e., the momentum) to each center parameter of the nucleons in the cluster, 
as done in Ref.~\cite{itagaki05}. 
The momentum direction is fixed to be orthogonal to the nucleon spin direction 
and it is opposite for spin-up and spin-down nucleons to obtain spin-orbit attraction. 
The $t$ cluster is located in the direction of $\boldsymbol{e}_z$ from the other cluster,
and the spins of one proton and neutron in the $t$ cluster are oriented to the $-\boldsymbol{e}_x$ 
direction while the spins of the other neutron is oriented in the $\boldsymbol{e}_x$ direction. 
Then, we choose the momentum direction to be $\boldsymbol{e}_y$, 
and we define the center parameters of the spin-up and spin-down nucleons in the $t$ cluster as 
$\boldsymbol{R}_{\uparrow / \downarrow} = d_t \boldsymbol{e}_z \pm i \lambda_t \boldsymbol{e}_y 
\ (d_t, \lambda_t ; {\rm real})$ 
in the rest frame of the $\alpha$ cluster. 
The center of mass of the $^6$He$+t$ or $\alpha+t$ cluster wave function is fixed to the origin. 
$\lambda_t$ characterizes the magnitude of the momentum of the nucleon in the $t$ cluster, or, 
in other words, the degree of $t$ cluster breaking. 
When $\lambda_t = 0$, the $t$ cluster has the ideal $(0s)^3$ configuration, 
while, when $\lambda_t$ is finite, the $t$ cluster is broken. 
In the case of $\lambda_t = d_t$ in the $d_t \rightarrow 0$ limit, 
one proton and two neutrons in the $t$ cluster occupy the H.O. $0p_{3/2}$ orbits around the $\alpha$ cluster. 
In this way, $t$ cluster breaking can be taken into account by changing the $\lambda_t$ value. 
It should be noted that the occupation probability of the single-particle orbits
(mainly the ratio between the $0p_{3/2}$ and $0p_{1/2}$ orbits) in the core is changed 
by introducing $t$ cluster breaking. 
We apply this method to the $t$ cluster 
in the $^6$He$+t$ cluster wave function and the $^9$Li DC wave function, 
and we consider $t$ cluster breaking as the change in core structure 
as well as $t$ cluster development.

\subsection{Projection and superposition of wave functions}
\label{sec:projection_superposition}

We describe $^9$Li 
by superposing the $^6$He+$t$ cluster wave functions and the $^9$Li DC wave functions. 
The $t$ cluster in each wave function can be broken 
using the method shown in Sec.~\ref{sec:t_breaking}. 
\begin{align}
\Psi^{J \pi}_M = 
\sum_K \Big( 
& \sum_i c_{Ki} \mathcal{P}^{J \pi}_{MK} 
\Phi_{^6{\rm He}+t}(\kappa, \lambda_t, d_t) \nonumber \\
+ & \sum_j c_{Kj} \mathcal{P}^{J \pi}_{MK} 
\Phi_{\rm DC}(\lambda_t,d_t,\beta,b_n)
\Big). 
\label{eq:Li9_wf}
\end{align}
The parameter $\lambda_t$, 
which characterizes the degree of $t$ cluster breaking, is added to each wave function
[Eqs.~(\ref{eq:He6+t_cluster_wf}) and (\ref{eq:DC_wf})]. 
$i$ and $j$ are the abbreviations of $i = \{ \kappa, \lambda_t,d_t \}$
and $j = \{ \lambda_t,d_t, \beta,b_n \}$. 
$\mathcal{P}^{J \pi}_{MK}$ is the projection operator 
to the eigenstate having the total spin-parity of $J^{\pi}$. 
In the present work, we consider only the ground state of $^9$Li with $J^{\pi} = 3/2^-$. 
The coefficients $c_{Ki}$ and $c_{Kj}$ in Eq.~(\ref{eq:Li9_wf}) 
are determined by diagonalizing the Hamiltonian in Sec.~\ref{sec:hamiltonian}. 

By superposing the $^6$He$+t$ cluster wave function, 
we describe the main structure of the $^9$Li ground state. 
This wave function describes the competition between the configurations of the two valence neutrons 
near the $\alpha+t$ core, that is, mainly, 
the spin-singlet configuration (dineutron formation) in the $p$ shell
and the $(0p_{3/2})^2$ configuration (dineutron dissociation due to the spin-orbit interaction). 
By superposing the $^9$Li DC wave function, 
we describe a dineutron whose size and spatial expansion from the core can change. 
We discuss how $t$ cluster breaking and development affect the dineutron component
near and far from the core. 
Although we do not take into account that 
the dineutron component far from the $\alpha+t$ core is dissociated due to the spin-orbit interaction, 
the $(0p_{3/2})^2$ configuration just at the surface is expected to be the dominant component 
of the dissociated dineutron for $^9$Li. 

We examine the adopted parameters in the superposition of Eq.~(\ref{eq:Li9_wf}). 
To clarify the effects of $t$ cluster breaking on dineutron correlation, we perform the calculations with and without $t$ cluster breaking. 
In the calculation with $t$ cluster breaking, 
we choose $\lambda_t = 0.0, 0.4$, and $0.8$ fm 
for both the $^6$He+$t$ wave functions and the $^9$Li DC wave functions. 
In the calculation without $t$ cluster breaking, 
we fix $\lambda_t$ at $0.0$ fm. 
In both calculations, 
the relative distance between the $^6$He or $\alpha$ cluster and the $t$ cluster is chosen as
$d_t = 1, 2, \ldots, 5$ fm for each value of $\lambda_t$ 
to consider $t$ cluster development. 
By comparing the results obtained from the calculations with and without $t$ cluster breaking, 
we can evaluate the effects of $t$ cluster breaking. 

In addition to the parameters $(\lambda_t, d_t)$ characterizing the core structure, 
each wave function has parameters 
characterizing the behavior of the two valence neutrons around the core. 
For the $^6$He$+t$ cluster wave function, the parameter $\kappa$ 
labels the $(0p)^2$ configurations of the two valence neutrons around the $\alpha$ cluster. 
We perform configuration mixing by superposing the wave functions with $\kappa$ as done in Ref.~\cite{enyo12}. 
The $^9$Li DC wave function has two parameters, $b_n$ and $\beta$, 
characterizing the dineutron size and the spatial expansion of the dineutron distribution from the core. 
We superpose $\beta = 2, 3, \ldots, 6$ fm 
and four $b_n$ values are determined for each $\beta$ as  
\begin{equation}
b_n = 1.5 \times \left( \frac{\beta-0.2}{1.5} \right)^{(i-1)/(4-1)} \ ( i = 1, \ldots, 4).
\label{eq:b_n}
\end{equation} 
These parameter choices are sufficient to describe the compact dineutron 
about the $\alpha+t$ core in the rather compact ground state of $^9$Li. 

Next, we examine the framework used to describe the other nuclei. 
In the case of $^{10}$Be, we consider dineutron features around the $\alpha+\alpha$ core 
by superposing the $^6$He$+\alpha$ cluster wave functions 
and the $^{10}$Be DC wave functions. 
Also, for $^{10}$Be, we take into account cluster breaking of one of the $\alpha$ clusters. 
The adopted parameters in the superposed wave functions are the same as those described for the $^9$Li case. 
We project the superposed wave functions to $J^{\pi} = 0^+$ for the $^{10}$Be ground state. 
As for the mirror nuclei, $^9$C and $^{10}$C, 
we merely exchange protons and neutrons in the framework adopted for $^9$Li and $^{10}$Be, respectively.

\subsection{Measurement of the dinucleon component}
\label{sec:measure_dineutron}

The aim of this work is to investigate the dinucleon components of $^9$Li, $^{10}$Be, and $^{9,10}$C, 
and to discuss their possible dinucleon features.
We define two measurements of the dineutron component around the core for each purpose. 
We mainly describe their definitions for the case of $^9$Li 
and add definitions as necessary for the other nuclei afterwards. 

The first measurement is the overlap with a DC wave function. 
The purpose of this overlap is to evaluate the dineutron size and expansion around a certain core in the state 
\begin{equation}
\mathcal{N}_{\rm DC}(\beta, b_n) = |\langle \mathcal{P}^{3/2^-}_{MK'} \Phi_{\rm DC}(\beta, b_n)|
\Psi^{3/2^-_1}_M \rangle|^2, 
\label{eq:overlap_DC}
\end{equation}
where $\Psi^{3/2^-_1}_M$ is the ground state wave function of $^9$Li calculated as Eq.~(\ref{eq:Li9_wf}). 
$\Phi_{\rm DC}$ is the test DC wave function whose core is $\alpha+t$ 
and has two characteristic parameters: 
the dineutron size, $b_n$, and its expansion from the core, $\beta$. 
We prepare a certain $\alpha+t$ core wave function for the test DC wave function. 
In the present work, the $\alpha+t$ core structure is characterized by two parameters: 
the degree of $t$ breaking, $\lambda_t$, and the $\alpha$-$t$ distance, $d_t$. 
We fix the parameters $(\lambda_t, d_t)$ to optimum values, change the parameters $(\beta,b_n)$ to various values, 
and observe how much the dineutron probability depends on the dineutron size and expansion about the specified core. 
We use this measurement in Sec.~\ref{sec:diproton}
to investigate the degree of dinucleon-size change for $^9$Li and $^9$C. 

The second measurement determines the compact dineutron probability 
at a certain distance around the $\alpha+t$ core 
with consideration for the degree of core cluster development. 
For the test wave function, we fix the dineutron size to a certain value 
and specify the distance between the compact dineutron and the $\alpha+t$ core 
as well as the distance between the $\alpha$ and $t$ clusters in the core. 
The overlap for the second measurement is written as follows. 
\begin{equation}
\mathcal{N}_{\rm dineutron}(d_{2n}, d_t)
= |\langle \Phi_{\rm dineutron} (d_{2n}, d_t) | \Psi^{3/2^-_1}_M \rangle|^2, 
\label{eq:overlap_dineutron}
\end{equation}
where $\Phi_{\rm dineutron}$ is the wave function 
containing the compact dineutron of a fixed size ($\nu = 0.22$ fm$^{-2}$) 
spherically distributed around the $\alpha+t$ core at a certain distance. 
The distance between the dineutron and the $\alpha+t$ core is specified by $d_{2n}$, 
and that between the $\alpha$ and $t$ clusters in the core is specified by $d_t$. 
For the overlap of $^9$Li with $t$ cluster breaking, 
we take into account $t$ cluster breaking in the $\alpha+t$ core 
of the test wave function as well. 
The details of these overlaps, $\mathcal{N}_{\rm dineutron}$, are described in the appendix. 
In the present work, 
we use this measurement to evaluate core structure changes on the dineutron in $^9$Li, 
and to compare dinucleon features in $^9$Li, $^{10}$Be, and $^{9,10}$C. 

\begin{figure}
\includegraphics[scale=0.25]{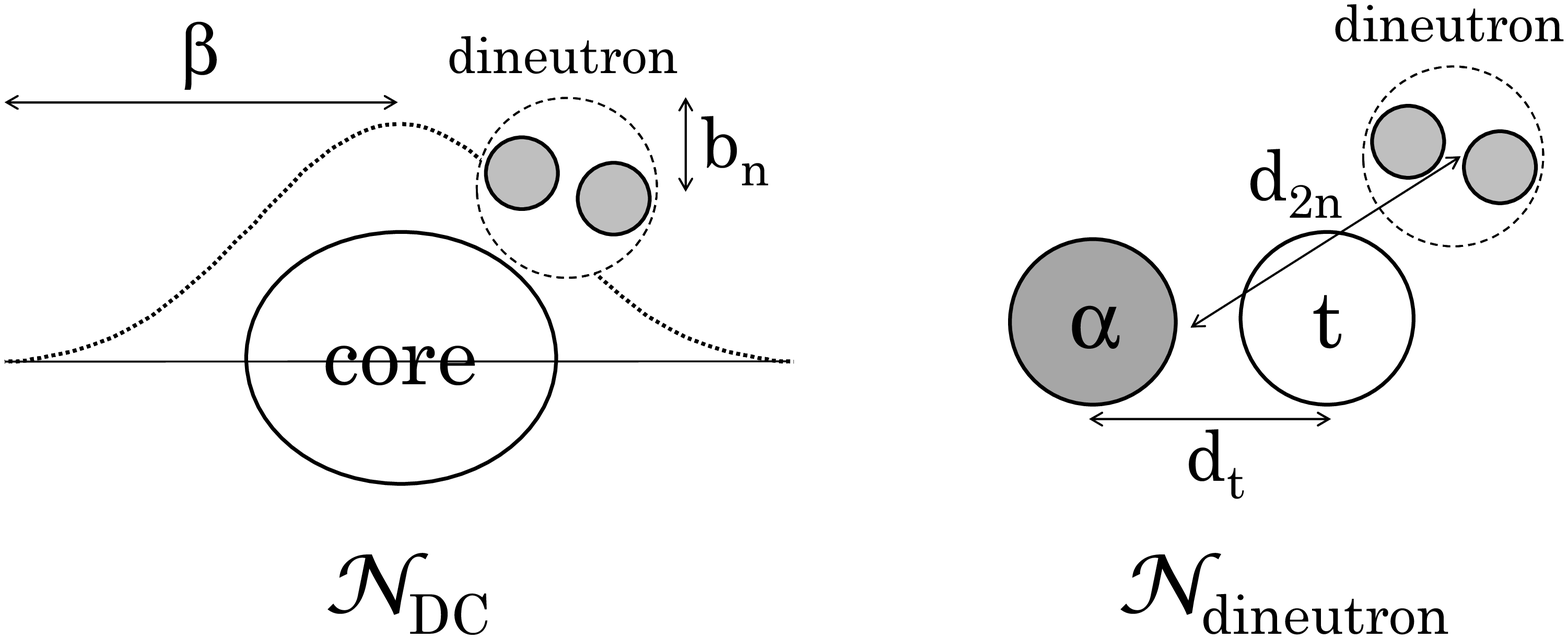}
\caption{Schematic representation of the two types of wave functions used for the $\mathcal{N}_{\rm DC}$ (left) 
and $\mathcal{N}_{\rm dineutron}$ (right) measurements. 
In $\mathcal{N}_{\rm DC}$, the key parameters are 
$b_n$ (the dineutron size) and $\beta$ (the dineutron expansion from the core). 
In $\mathcal{N}_{\rm dineutron}$, the key parameters are $d_{2n}$ (the core-dineutron distance) 
and $d_t$ (the $\alpha$-$t$ distance). }
\label{fig:N_dineutron}
\end{figure}

Schematic representations of the two types of test wave functions of each overlap are shown in Fig.~\ref{fig:N_dineutron}. 
The overlap involving the DC wave function, $\mathcal{N}_{\rm DC}$ [Eq.~(\ref{eq:overlap_DC})], is a measurement 
of the dineutron size, $b_n$, and its expansion, $\beta$, from the core. 
It is noted here that the expansion, $\beta$, is not a distance 
but a degree of spatial expansion from the core. 
On the other hand, the second measurement, $\mathcal{N}_{\rm dineutron}$ [Eq.~(\ref{eq:overlap_dineutron})], 
evaluates the core-dineutron distance, $d_{2n}$, 
and the $\alpha$-$t$ distance in the core, $d_t$, 
for a dineutron size fixed at a certain value. 

The compact dineutron probability, $\mathcal{N}_{\rm dineutron}$, of the $^{10}$Be ground state is
defined similarly to Eq.~(\ref{eq:overlap_dineutron}). 
For $^{10}$Be, the core is $\alpha+\alpha$. 
Measurement of the diproton probability for $^{9,10}$C 
is conducted the same as that for $^9$Li and $^{10}$Be
except that protons and neutrons are exchanged. 

\section{Dineutron correlation in $^9$Li and $^{10}$Be}
\label{sec:dineutron}

In this section, we discuss 
the degree of dineutron formation in the inner and surface regions 
and the degree of dineutron expansion in the outer region in the ground states of $^9$Li and $^{10}$Be. 
Comparing the results with and without $t$ cluster breaking, 
we clarify the effect of core structure changes on the dineutron features of $^9$Li. 
We also discuss the interaction dependence of the dineutron probability. 
Additionally, we compare the dineutron probability of $^9$Li with that of $^{10}$Be 
and discuss the differences in the dineutron features
arising from core structure differences.

\subsection{Hamiltonian}
\label{sec:hamiltonian}

In the present work, we use the Hamiltonian 
\begin{equation}
H = T - T_G + V_{\rm cent} + V_{\rm LS} + V_{\rm Coul}, 
\label{eq:hamiltonian}
\end{equation}
where $T$ and $T_G$ are the total and center-of-mass kinetic energies. 
Because all of the Gaussian widths of the DC wave function are not equal, 
the center-of-mass motion cannot be removed exactly. 
We therefore treat the center-of-mass motion approximately 
by extracting the expectation value of the center-of-mass kinetic energy, $T_G$, 
from the total Hamiltonian. 
$V_{\rm Coul}$ is the Coulomb force 
that is approximated by the summation of seven Gaussian functions. 
$V_{\rm cent}$ and $V_{\rm LS}$ are the effective central and spin-orbit interactions 
for which we use the Volkov No.2 force \cite{volkov65} 
and the spin-orbit part of the G3RS force \cite{tamagaki68}, respectively. 
In this work, we choose 
the Majorana ($m$) parameter $m = 0.58$, 
the Bartlett ($b$) and Heisenberg ($h$) parameters $b=h=0.0$ in $V_{\rm cent}$, 
and the strength of $v_{\rm LS} = 1600$ MeV in $V_{\rm LS}$. 
This $v_{\rm LS}$ value is the same as that in the preceding study of $^9$Li \cite{enyo12}.
The parameters of the central force are chosen 
so as to reproduce the $2n$ separation energy in $^9$Li shown 
in Table~\ref{tab:energy_radius}. 
In the calculation for $^9$Li, we also use two other parameter sets 
to see the effect of spin-orbit interaction on dineutron formation, as described later. 
Here, we should comment about the choice of $b=h=0.0$ in the central force. 
Two neutrons in the spin-singlet channel are bound in accordance with these values, 
contradictory to experimental fact. 
Later, we show that our choices for parameters $b$ and $h$  barely change the qualitative features of dineutron correlation 
by comparing calculations using $b=h=0.0$
with $b=h=0.125$, which reproduces the $n$-$n$ unbound feature.

\subsection{Ground state energies and radii}
\label{sec:energy_radius}

\begin{table*}
\caption{The total energy, the two-nucleon separation energy ($S_{2n, 2p}$),
and the matter, proton and neutron radii ($r_m$, $r_p$, and $r_n$)
of the $^9$Li, $^{10}$Be, and $^{9,10}$C ground states. 
The experimental values of the matter radii have been referred from those in Ref.~\cite{ozawa01}.}
\label{tab:energy_radius}
\begin{ruledtabular}
\begin{tabular}{lllccccccc}
&&& Energy (MeV) & $S_{2n, 2p}$ (MeV) &
$r_m$ (fm) & $r_p$ (fm) & $r_n$ (fm) & \\ 
\hline
$^9$Li 
& w/o $t$ breaking && $-41.69$ & 5.07 & 2.42 & 2.23 & 2.51 & \\
& with $t$ breaking && $-42.85$ & 5.62 & 2.37 & 2.18 & 2.46 & \\
& Expt. && $-45.34$ & 6.09 & $2.32 \pm 0.02$ &&& \\
\hline
$^{10}$Be 
&  w/o $\alpha$ breaking && $-63.46$ & 8.33 & 2.46 & 2.38 & 2.52 & \\
& with $\alpha$ breaking && $-65.14$ & 9.43 & 2.41 & 2.32 & 2.48 & \\
& Expt. && $-64.98$ & 8.38 & $2.30 \pm 0.02$ &&& \\
\hline
$^9$C
& w/o $h$ breaking && $-35.86$ & 0.76 & 2.45 & 2.55 & 2.24 & \\
& with $h$ breaking && $-36.73$ & 1.04 & 2.42 & 2.52 & 2.21 & \\
& Expt. && $-39.04$ & 1.44 & $2.42 \pm 0.03$ &&& \\
\hline
$^{10}$C 
& w/o $\alpha$ breaking && $-59.06$ & 3.93 & 2.49 & 2.55 & 2.40 & \\
& with $\alpha$ breaking && $-60.76$ & 5.05 & 2.46 & 2.53 & 2.35 & \\
& Expt. && $-60.32$ & 3.73 & $2.27\pm 0.03$ &&&
\end{tabular}
\end{ruledtabular}
\end{table*}

The energies and radii of $^9$Li, $^{10}$Be, and $^{9,10}$C ground states are shown in Table~\ref{tab:energy_radius}. 
We calculate the $2n$ separation energy, $S_{2n}$, for $^9$Li 
by extracting the energy of the $^7$Li core as
\begin{equation}
S_{2n} \equiv - \big( \langle \Psi^{3/2^-_1}_M | H | \Psi^{3/2^-_1}_M \rangle
- \langle \Psi_{\alpha+t} |H| \Psi_{\alpha+t} \rangle \big), 
\label{eq:2n_separation_energy}
\end{equation}
where $\Psi^{3/2^-_1}_M$ is the ground state wave function of $^9$Li [Eq.~(\ref{eq:Li9_wf})]. 
$\Psi_{\alpha+t}$ is the ground state (the first $3/2^-$ state) of $^7$Li 
described by the superposition of the $\alpha+t$ cluster wave functions 
with or without $t$ cluster breaking. 
The adopted parameters in the basis wave functions for $^7$Li,
$\lambda_t$ for the degree of $t$ cluster breaking and $d_t$ for the $\alpha$-$t$ distance, 
are the same as those found in the $^6$He$+t$ cluster wave function
used to describe $^9$Li. 
In a similar manner as for $^{10}$Be and $^{9,10}$C, 
we evaluate the $2n$ and $2p$ separation energies, $S_{2n}$ and $S_{2p}$, 
by extracting the $\alpha+\alpha$, $\alpha+h$, or $\alpha+\alpha$ core energy, respectively. 
The $2N$ separation energies for $A=9$ nuclei, $S_{2n}$ of $^9$Li and $S_{2p}$ of $^9$C, 
calculated with $t$ or $h$ cluster breaking,
are reproduced as shown in Table~\ref{tab:energy_radius}. 
For $^{10}$Be and $^{10}$C, 
the two valence nucleons are slightly overbound in comparison with the experimental values. 
As yet, it has not proved possible to find an interaction parameter set 
that reproduces all the data of $S_{2N}$ for both $A=9$ and $A=10$ nuclei simultaneously
for the presently employed two-body effective interactions. 
We have determined that such overbinding is not crucial for the present results regarding $A=10$ nuclei 
by adjusting the interaction parameters to reproduce $S_{2N}$ in the $A=10$ nuclei. 
The matter radii of $^9$Li, $^{10}$Be, and $^{10}$C are consistent with the experimental values,
except that of $^{10}$C is overestimated. 
For calculations with core cluster breaking, 
the core is observed to shrink for all nuclei, 
which is demonstrated by the smaller values of $r_p$ for $^9$Li and $^{10}$Be, and by $r_n$ for $^{9,10}$C, 
compared with those given by calculations without core cluster breaking. 
As a result, the mean-field potential formed by the core becomes deeper 
and $S_{2N}$ becomes larger. 

In the following sections, we first describe the dineutron probability in $^9$Li,  
and compare the results with and without $t$ cluster breaking in the core 
to evaluate the effect of core structure changes on dineutron features.  
Next, we compare the dineutron features of $^{10}$Be and $^9$Li. 
We also investigate the diproton correlation of $^{9,10}$C 
and discuss the differences in the dinucleon correlation 
between $^9$Li and $^9$C, and between $^{10}$Be and $^{10}$C in Sec.~\ref{sec:diproton}.

\subsection{Effect of core structure changes on the dineutron correlation in $^9$Li}
\label{sec:dineutron_Li9}

\begin{figure}[b]
\includegraphics[scale=0.65]{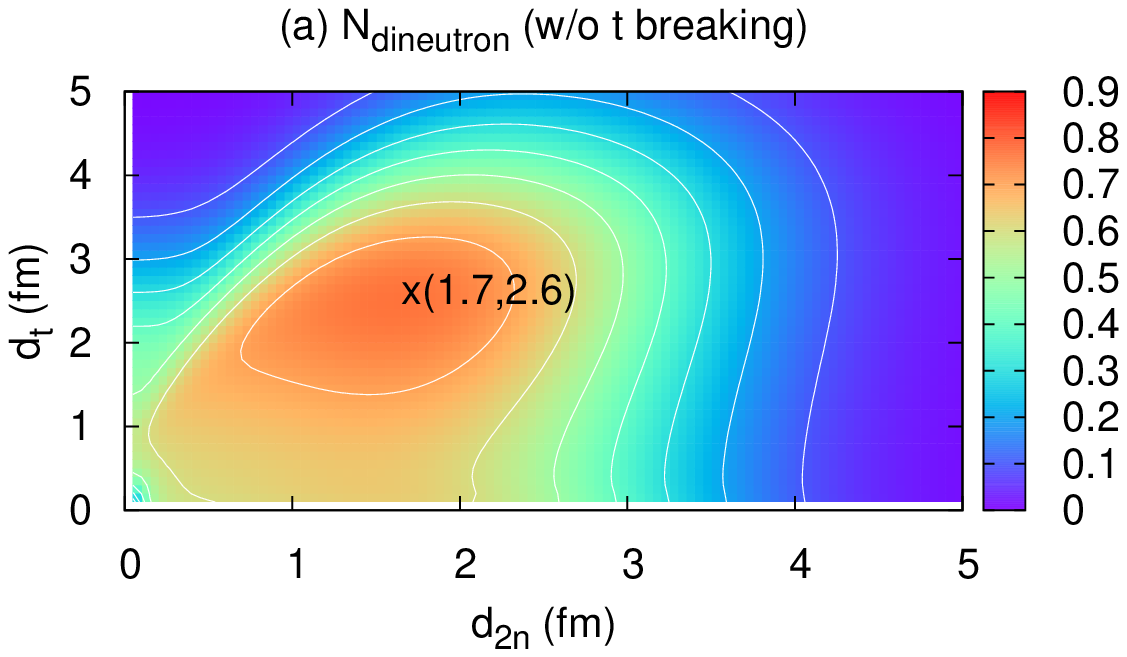} \\
\includegraphics[scale=0.65]{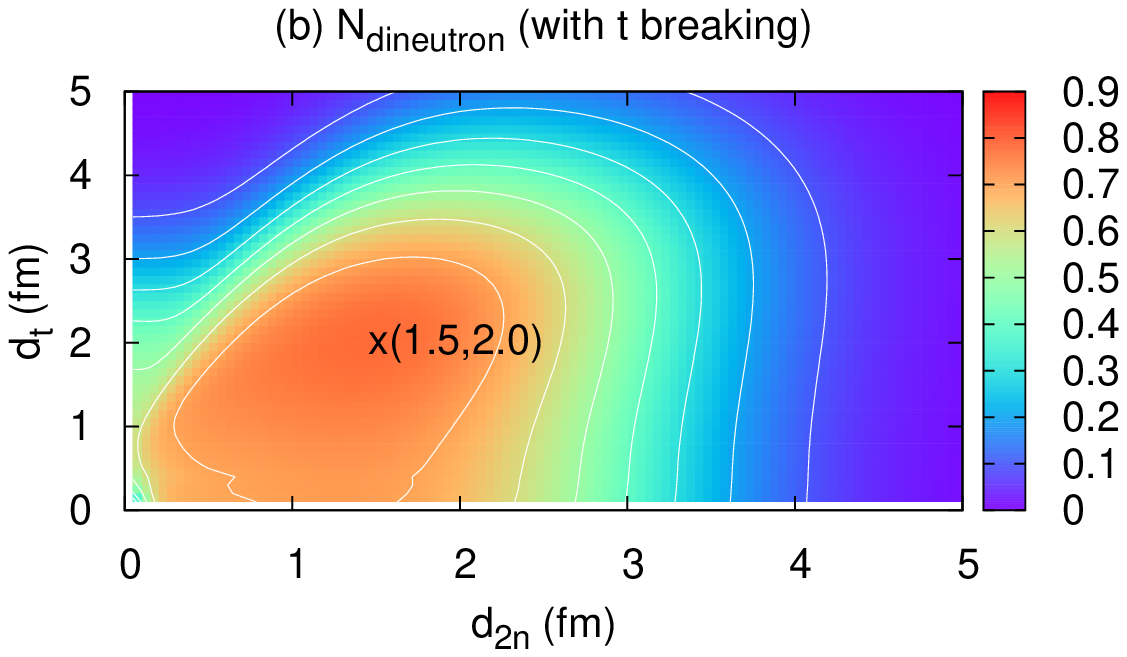}
\caption{(Color online) The dineutron probability, $\mathcal{N}_{\rm dineutron}$, for the ground state of $^9$Li
on the $d_{2n}$-$d_t$ plane. 
The upper figure (a) is the result without cluster $t$ breaking, 
and the lower figure (b) is the result with $t$ cluster breaking.
The symbol x represents the peak position. }
\label{fig:Li9_3D_dineutron}
\end{figure}

To investigate dineutron formation and spatial expansion from the $\alpha+t$ core 
in the $^9$Li ground state, 
we examine the probability, $\mathcal{N}_{\rm dineutron}$, for $^9$Li
as a function of $(d_{2n}, d_t)$  plotted in Figs.~\ref{fig:Li9_3D_dineutron}(a) without $t$ cluster breaking, and (b) with $t$ cluster breaking. 
$d_{2n}$ corresponds to the core-dineutron distance 
and $d_t$ corresponds to the $\alpha$-$t$ distance.
In the calculation involving  $t$ cluster breaking, 
the probability is enhanced in the small $d_t$ region ($d_t \lesssim 2$ fm), 
indicating that the $\alpha+t$ core is shrunk. 
As a result of core shrinkage, 
the dineutron is pulled nearer the core due to the strong binding between the core and the valence neutrons, 
and the peak value of $d_{2n}$ is correspondingly diminished. 

\begin{figure}[b]
\includegraphics[scale=0.6]{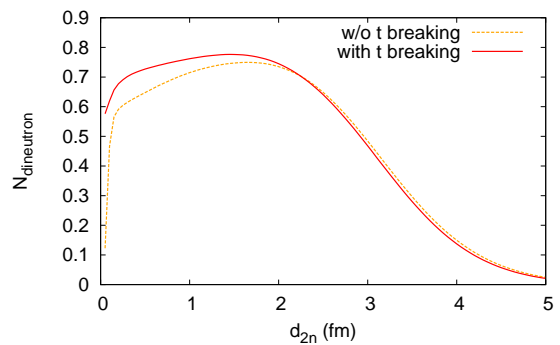} 
\caption{(Color online) The dineutron probability, $\mathcal{N}_{\rm dineutron}$,
as a function of the dineutron-core distance, $d_{2n}$, in the ground state of $^9$Li 
calculated with (solid line) and without (dashed line) $t$ cluster breaking. }
\label{fig:Li9_dineutron_S}
\end{figure}

We focus on the degree of dineutron formation and the spatial expansion from the core
under conditions where the $t$ cluster can or cannot be broken. 
To examine these dineutron features in more detail, 
we plot the dineutron probability as a function of the core-dineutron distance, $d_{2n}$, 
in Fig.~\ref{fig:Li9_dineutron_S}. 
Here, we take the maximum amplitude of $\mathcal{N}_{\rm dineutron}(d_{2n}, d_t)$ 
in Fig.~\ref{fig:Li9_3D_dineutron}
for each $d_{2n}$ value. 
As seen in the figure, the change in the dineutron probability due to $t$ cluster breaking is minor
in the outer region of $d_{2n} \gtrsim 2.5$ fm. 
On the other hand, the dineutron probability in the calculation for the inner region of $d_{2n} \lesssim 2.0$ fm 
is enhanced with $t$ cluster breaking. 
The key to such enhancement of the inner dineutron component 
is the change in the occupation probability of the $0p_{3/2}$ neutron orbits in the $^7$Li core. 
For simplicity, we assume that the $t$ cluster and two valence neutrons 
are distributed around the $\alpha$ cluster at a small distance. 
When the dineutron is distributed near the core, 
two neutrons in the dineutron feel the spin-orbit interaction from the core 
and the dineutron possibly dissociates to two independent neutrons in the major shell. 
Namely, the dineutron dissociation here results from competition 
between the dineutron (spin-singlet) component and the $(0p_{3/2})^2$ component of two neutrons 
around the core. 
In the situation where the $t$ cluster can be broken, 
the two neutrons in the $t$ cluster tend to occupy the $0p_{3/2}$ orbits to gain the spin-orbit attraction, 
and the $0p_{3/2}$ orbits are partially forbidden for the two valence neutrons around the $^7$Li core. 
Consequently, the two valence neutrons cannot occupy the $0p_{3/2}$ orbits predominantly
so that dineutron dissociation to the $(0p_{3/2})^2$ component is suppressed
and the dineutron survives even in the inner region. 
As a result, the dineutron component is enhanced
in the situation where the Pauli principle between the neutrons in the dineutron 
and those in the $t$ cluster becomes significant near the $\alpha$ cluster
(i.e., the region of small $d_{2n}$ and $d_{\lambda}$ shown in Fig.~\ref{fig:Li9_3D_dineutron}). 
Accordingly, we conclude that core cluster breaking, 
i.e., the occupation probability of the lower of the $ls$-splitting orbits in the core
($0p_{3/2}$ in the $^9$Li case), 
significantly affects the degree of dineutron formation.

\subsection{Interaction dependence of the dineutron correlation of $^9$Li}
\label{sec:dineutron_Li9_interaction}

\begin{figure}[t]
\includegraphics[scale=0.6]{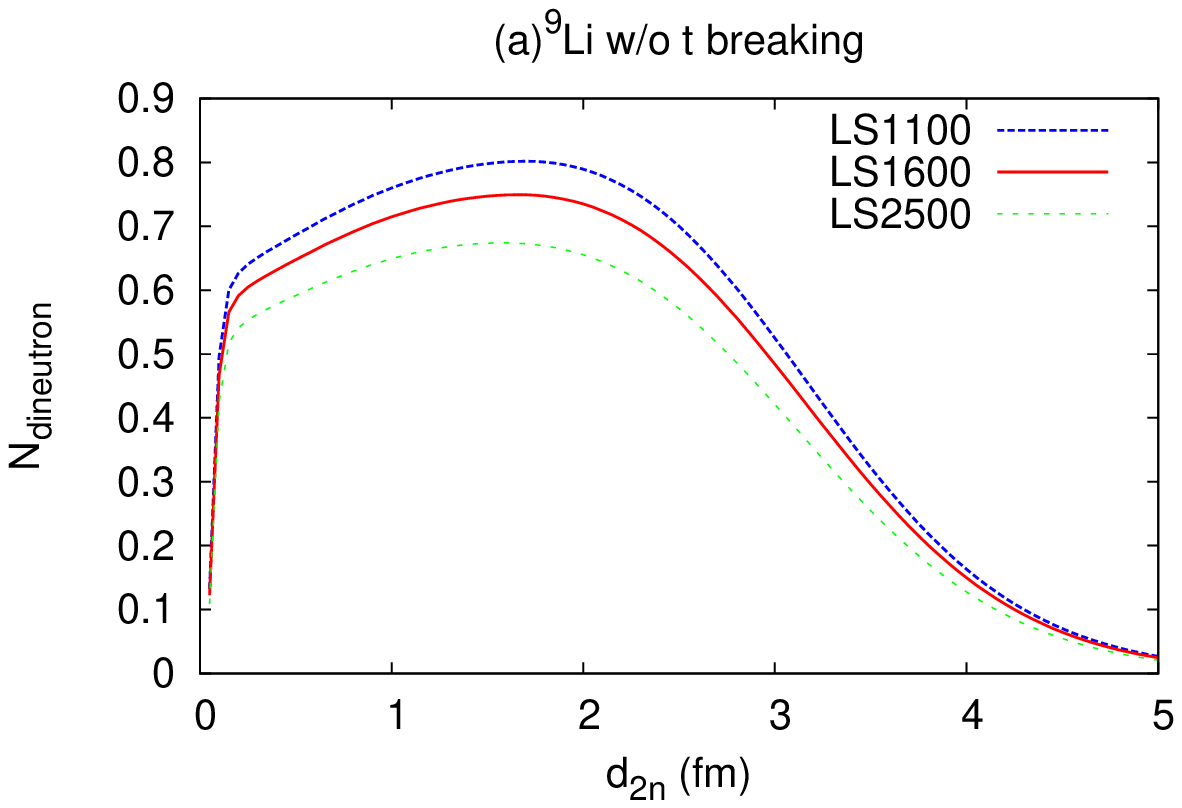} \\
\includegraphics[scale=0.6]{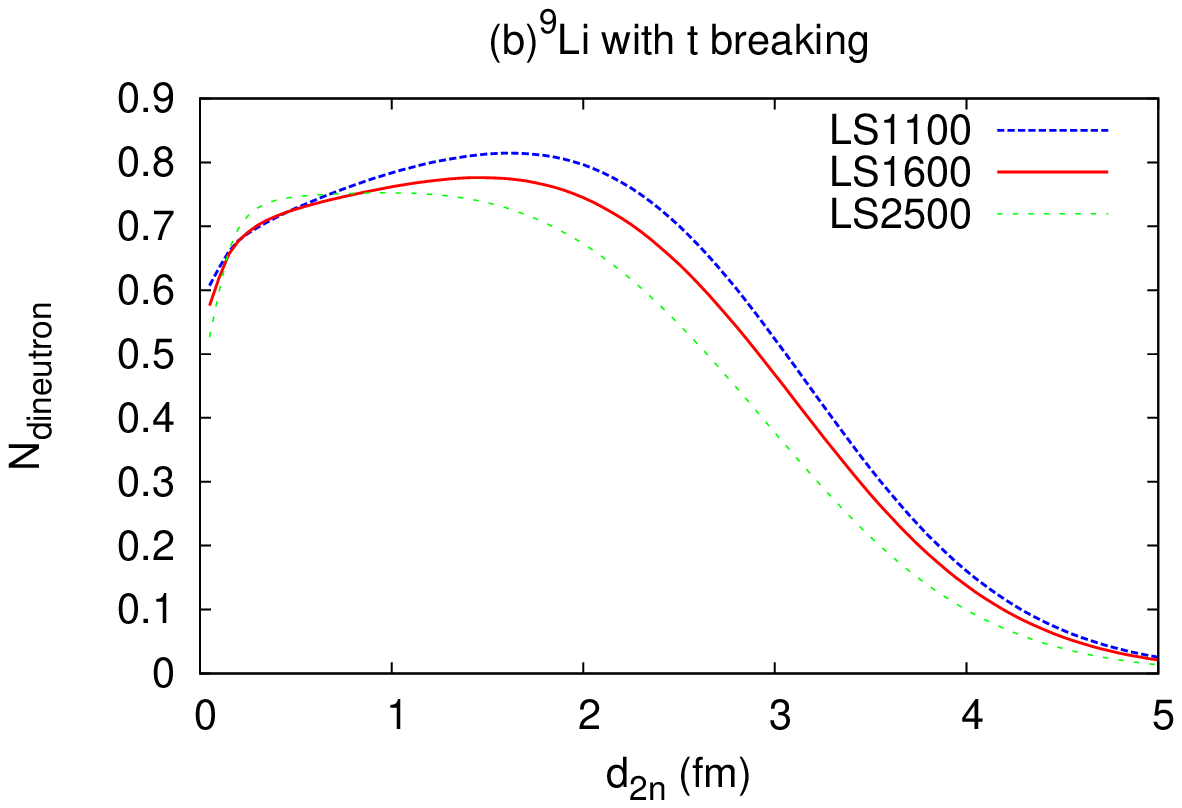} 
\caption{(Color online) The dineutron probability, $\mathcal{N}_{\rm dineutron}$, of the ground state of $^9$Li 
(a) without $t$ cluster breaking and (b) with the $t$ cluster breaking, 
calculated with LS1100 (dashed line), LS1600 (solid line) and LS2500 (dotted line). }
\label{fig:Li9_dineutron_LS}
\end{figure}

We show the effects of spin-orbit interaction on dineutron probability 
by using three interaction parameter sets 
having different strengths of the spin-orbit interaction, $v_{\rm LS}$. 
The Majorana parameter, $m$, in the central force is adjusted 
to give almost the same $2n$ separation energy, $S_{2n}$,  
in the calculation with $t$ cluster breaking. 
The first set has $m=0.58, \ b=h=0.0$, and $v_{\rm LS} = 1600$ MeV, as used above. 
For the second set, where the spin-orbit interaction is weaker, we choose $m=0.56, \ b=h=0.0$, and $v_{\rm LS} = 1100$ MeV. 
In the third, where the spin-orbit interaction is stronger, we choose $m=0.62, \ b=h=0.0$ and $v_{\rm LS} = 2500$ MeV. 
We label these parameter sets as ``LS1600'', ``LS1100'', and ``LS2500'', respectively. 
The ground state energy with $t$ cluster breaking is
$-43.38$ MeV ($S_{2n} = 5.66$ MeV) for parameter set LS1100,
$-42.24$ MeV ($S_{2n} = 5.87$ MeV) for LS2500, and 
$-42.85$ MeV ($S_{2n} = 5.62$ MeV) for LS1600. 
The dineutron probability, $\mathcal{N}_{\rm dineutron}$, 
for the states with and without $t$ cluster breaking,
calculated with these parameter sets is plotted as a function of $d_{2n}$ in Fig.~\ref{fig:Li9_dineutron_LS}. 
The dineutron probability over the entire region depends on the strength of the spin-orbit interaction 
in the calculations with and without $t$ cluster breaking. 
These data clearly reflect that a dineutron is readily dissociated to gain the spin-orbit energy. 

In the calculation with $t$ cluster breaking, 
the dineutron probability inner region ($d_{2n} \lesssim 1.5$ fm)
is enhanced when the spin-orbit interaction is strong. 
This is because, when the core cluster is largely broken due to strong spin-orbit interaction, 
the dineutron enhancement mechanism 
due to the partial blocking of the $0p_{3/2}$ orbits
mentioned in Sec.~\ref{sec:dineutron_Li9} is strongly operational. 

\begin{figure}[t]
\includegraphics[scale=0.6]{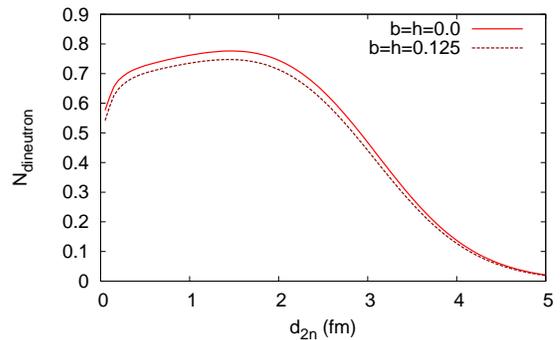} 
\caption{(Color online) The dineutron probability, $\mathcal{N}_{\rm dineutron}$, 
in the ground state of $^9$Li with $t$ cluster breaking 
calculated with $m=0.58, v_{\rm LS} = 1600$ MeV, and   $b=h=0$ (dashed line)  and $b=h=0.125$ (solid line). }
\label{fig:Li9_dineutron_bh}
\end{figure}

Next, we show the dependence of the dineutron correlation 
on the strength of the $n$-$n$ interaction. 
We compare the dineutron probability, $\mathcal{N}_{\rm dineutron}$, 
in the ground state with $t$ cluster breaking 
calculated with the original parameter set of $m=0.58, \ b=h=0.0$, and $v_{\rm LS} = 1600$ MeV  
in addition to being calculated by adjusting the Bartlett and Heisenberg parameters in the central force to $b=h=0.125$, 
which reproduces the unbounded feature of the two neutrons in the spin-singlet channel. 
With $b=h=0.125$, 
the ground state energy of $^9$Li is $-40.78$ MeV ($S_{2n} = 3.54$ MeV), 
and the two neutrons are underbound to the core,
relative to the calculation with $b=h=0.0$, 
because of the weaker $n$-$n$ interaction. 
We plot their dineutron probability in Fig.~\ref{fig:Li9_dineutron_bh}. 
In the state calculated with $b=h=0.0$, 
the dineutron probability is certainly larger because of the stronger $n$-$n$ interaction. 
However, the amplitudes are not so different 
and the dineutron features seem to be almost the same qualitatively in these calculations.

\subsection{Comparison the dineutron correlation of $^{10}$Be with that of $^9$Li}
\label{sec:dineutron_Be10}

\begin{figure}[b]
\includegraphics[scale=0.6]{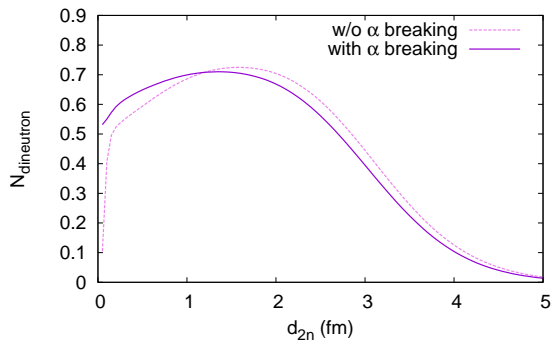} 
\caption{(Color online) The dineutron probability, $\mathcal{N}_{\rm dineutron}$ in the ground state of $^{10}$Be
calculated with (solid line) and without (dashed line) $\alpha$ cluster breaking . }
\label{fig:Be10_dineutron_S}
\end{figure}

Here, we consider the dineutron probability in the $^{10}$Be ground state
and compare it with that of $^9$Li. 
We also examine the effect of core structure change on the dineutron probability of $^{10}$Be, 
and clarify the differences between dineutron features in $^{10}$Be and $^9$Li. 

Fig.~\ref{fig:Be10_dineutron_S} shows the dineutron probability, $\mathcal{N}_{\rm dineutron}$,   
as a function of the distance between the $\alpha+\alpha$ core and a compact dineutron, $d_{2n}$ 
. 
For $^{10}$Be, the change in the probability due to core cluster breaking 
is quite different from that of $^9$Li [Fig.~\ref{fig:Li9_dineutron_S}]. 
The inner dineutron enhancement ($d_{2n} \lesssim 1.5$ fm) is not significant 
because the $\alpha$ cluster in the $\alpha+\alpha$ core is rigid and barely broken 
so that the enhancement mechanism due to the core cluster breaking in $^9$Li
mentioned in Sec.~\ref{sec:dineutron_Li9} is not so applicable.  
On the other hand, the dineutron probability 
in the outer region ($d_{2n} \gtrsim 2$ fm) decreases considerably. 
This is because the $n$-$\alpha$ attraction is stronger than the $n$-$t$ attraction,
and the $\alpha+\alpha$ core shrinkage, 
though it is not as notable as the $\alpha+t$ core shrinkage, 
markedly affects dineutron expansion from the core. 
Moreover, the spin-orbit interaction from the $\alpha$ cluster is stronger 
and dissociates the dineutron more readily compared with the $t$ cluster.

\begin{figure}
\includegraphics[scale=0.6]{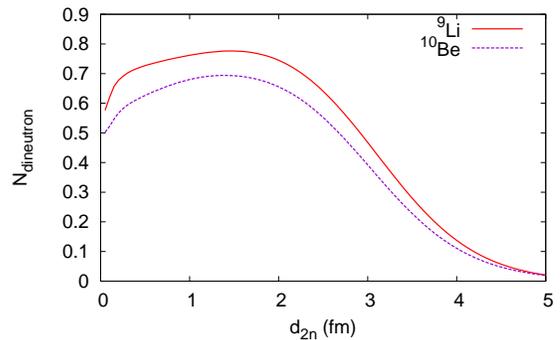} 
\caption{(Color online) The dineutron probability, $\mathcal{N}_{\rm dineutron}$, in the ground states of $^9$Li (solid line) 
and $^{10}$Be (dashed line) calculated with core cluster ($t$ or $\alpha$) breaking. }
\label{fig:N=6_dineutron}
\end{figure}

Finally, we compare the dineutron probability, $\mathcal{N}_{\rm dineutron}$, between $^9$Li and $^{10}$Be 
calculated with core cluster breaking, as shown in Fig.~\ref{fig:N=6_dineutron}. 
The dineutron probability of $^9$Li is larger than that in $^{10}$Be over the entire region. 
This is because the $n$-core attraction of $^9$Li is weaker than that of $^{10}$Be.
As a result, the dissociation effect due to the spin-orbit attraction on the dineutron is weaker, 
and, consequently, the dineutron can be readily formed in the inner and surface regions 
and distributed to the outer region from the core for $^9$Li. 
It can be concluded that both the dineutron spatial expansion from the core 
and the degree of dineutron formation strongly depend on the core structure.

\section{Diproton correlation in $^{9,10}$C}
\label{sec:diproton}

In Sec.~\ref{sec:dineutron} 
we discussed the compact dineutron probability in the ground states of $^9$Li and $^{10}$Be. 
In this section, we investigate diproton correlation of $^{9,10}$C, 
which are the mirror nuclei of $^9$Li and $^{10}$Be.  
To begin with, we compare the diproton probability of $^{9,10}$C and 
the dineutron probability of $^9$Li and $^{10}$Be 
and show that there are no qualitative differences between them. 
Next, we discuss the quantitative differences between dineutron and diproton correlation in more detail. 
We investigate the dinucleon-size changes of $^9$Li and $^9$C, 
and clarify the Coulomb effects of the dinucleon features.

\subsection{Comparison between the dinucleon correlation of $^9$Li, $^{10}$Be, and $^{9,10}$C}
\label{sec:dinucleon}

First, we elaborate on the differences in structure and binding of the valence nucleons 
of $^9$Li, $^{10}$Be, and $^{9,10}$C, as
shown in Table~\ref{tab:energy_radius}.
In the following section, we discuss the results with core cluster breaking. 
For $^9$C, the two valence nucleons are bound more loosely than those for $^9$Li, 
which is indicated by the smaller value of $S_{2N}$ shown for $^9$C
($S_{2p} \sim 1.0$ MeV for $^9$C and  $S_{2n} \sim 5.6$ MeV for $^9$Li). 
Such loose binding originates in the Coulomb force, 
mainly between the $\alpha+h$ core and the valence protons. 
However, in spite of such loose core-$N$ binding, 
the spatial expansion of the valence protons of $^9$C is almost the same as
that of the valence neutrons of $^9$Li, 
which is seen by the proton radius of $^9$C ($r_p \sim 2.52$ fm) 
being not significantly different from the neutron radius of $^9$Li ($r_n \sim 2.46$ fm). 
This is because the Coulomb barrier formed by the core 
prevents the valence protons from spreading far from the core. 
Additionally, since the neutron radius of $^9$C ($r_n \sim 2.21$ fm) is as large as 
the proton radius of $^9$Li ($r_p \sim 2.18$ fm), 
we suppose that the $\alpha+h$ core structure of $^9$C is not so different 
from that of the $\alpha+t$ core of $^9$Li. 
Also for $^{10}$Be and $^{10}$C, 
the expectation values of the radii are similar to each other 
and there seems to be no difference in their structures. 
Therefore the Coulomb force does not change the fundamental structure of these nuclei, 
which are not extremely loosely bound systems. 

\begin{figure}[t]
\includegraphics[scale=0.6]{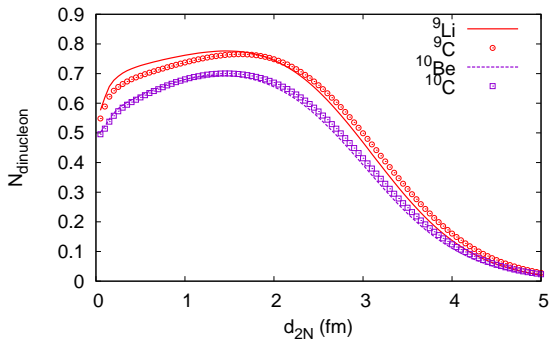} 
\caption{(Color online) The dinucleon probability, $\mathcal{N}_{\rm dinucleon}$, in the ground state of $^9$Li (solid line), 
$^{10}$Be, (dashed line), and $^{9,10}$C (open circle and square symbols, respectively)
calculated with core cluster breaking. }
\label{fig:mirror_dinucleon}
\end{figure}

Next, we consider the diproton probability, $\mathcal{N}_{\rm diproton}$, 
which is the component of a compact diproton at a distance of $d_{2p}$ from the core, 
defined in the same way as the dineutron probability of Eq.~(\ref{eq:overlap_dineutron}). 
The dinucleon probability, $\mathcal{N}_{\rm dinucleon}$, 
of $^9$Li, $^{10}$Be, and $^{9,10}$C is shown in Fig.~\ref{fig:mirror_dinucleon}. 
Comparing these results for $^9$C and $^9$Li,  
the difference in the dinucleon probability
is not so remarkable in spite of a rather large difference in $S_{2N}$ 
because the diproton expansion from the core is suppressed by the Coulomb barrier which the core forms. 
The differences between $^{10}$C and $^{10}$Be are smaller still. 
The Coulomb effect is surely minor for dinucleon correlation, 
but it should affect it quantitatively, more or less. 
In the following section, 
we clarify the quantitative differences between the diproton features of $^9$C and the dineutron features of $^9$Li. 

\subsection{Quantitative differences between dineutron and diproton features} 
\label{sec:dineutron_diproton}

\begin{figure}[b]
\includegraphics[scale=0.6]{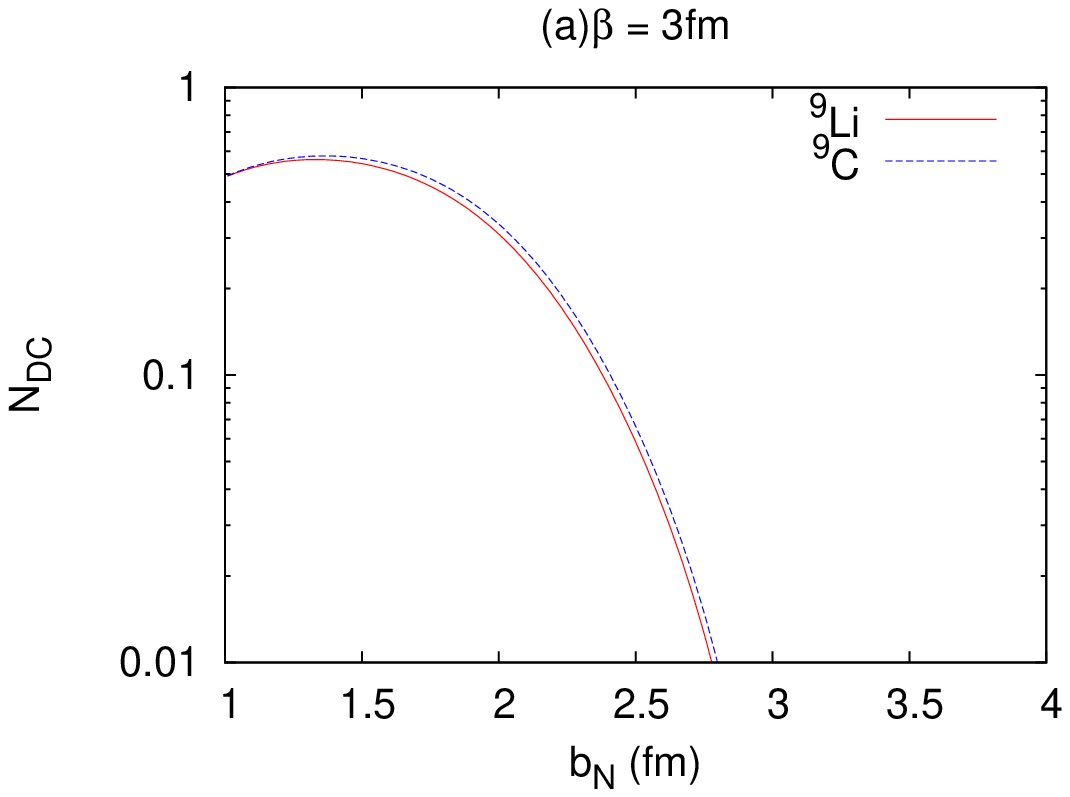} \\
\includegraphics[scale=0.6]{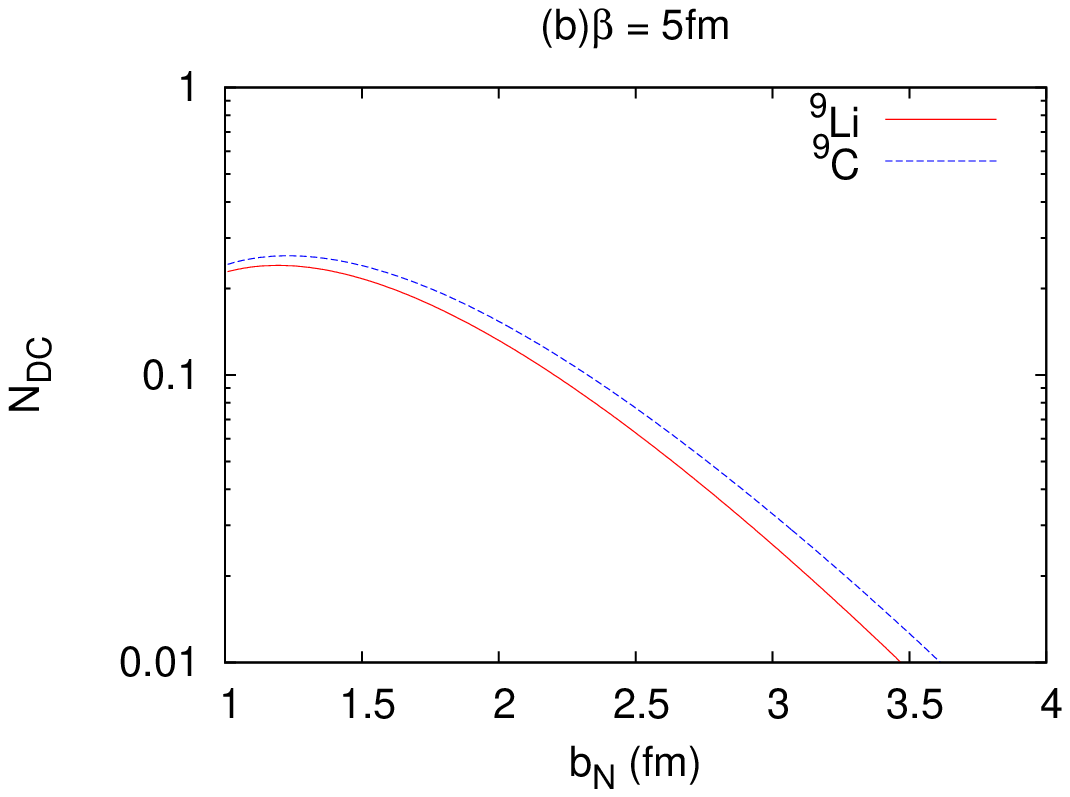}
\caption{(Color online) The dinucleon probability, $\mathcal{N}_{\rm DC}$, 
of $^9$Li (solid line) and $^9$C (dashed line) as a function of the dinucleon size, $b_N$, 
where the dinucleon expansion from the core, $\beta$, is fixed to (a) $3$ and (b) $5$ fm.}
\label{fig:overlap_b}
\end{figure}

In this subsection, 
we discuss the Coulomb effect on dinucleon features, 
especially the dinucleon size. 
We first investigate the Coulomb effect on the dinucleon size quantitatively 
by comparing the diproton in $^9$C with the dineutron in $^9$Li. 
Here, to consider the change in the dinucleon size, 
we show the overlap with the DC wave function, $\mathcal{N}_{\rm DC}$, 
depending on the dinucleon size and expansion from the core in the ground states of $^9$Li and $^9$C, 
defined in Eq.~(\ref{eq:overlap_DC}). 
The core of the test DC wave function used to measure the dineutron component
is the $\alpha+t$ or $\alpha+h$ cluster wave function 
described in Sec.~\ref{sec:DC_wf},  
which is characterized by the parameters $d_{t,h}$
(the relative distance between the $\alpha$ and the $t$ or $h$ clusters) 
and $\lambda_{t,h}$ (the degree of $t$ or $h$ cluster breaking). 
We fix $(\lambda_{t,h}, d_{t,h}) = (0.4, 2.0)$ 
so as to provide a sufficiently large overlap with the ground state of $^9$Li or $^9$C, 
and this set is reasonable for the $\alpha+t$ core of $^9$Li or the $\alpha+h$ core of $^9$C. 
We calculate the overlap with $K'=+1/2$ for the test DC wave function, 
which is the dominant component. 
By allowing $b_N$ to vary under a certain fixed value of $\beta$ in the test DC wave function, 
we observe a change in the dinucleon size ($b_N$) depending on the distribution around the core ($\beta$) 
in the ground states of $^9$Li and $^9$C. 
We plot the dinucleon probability, $\mathcal{N}_{\rm DC}$, as a function of the dinucleon size, $b_N$, 
fixing the dinucleon expansion from the core, $\beta$, to $3$ and $5$ fm in Fig.~\ref{fig:overlap_b}. 
When the dinucleon is near the core ($\beta = 3$ fm), 
both the dinucleon probability amplitudes of $^9$Li and $^9$C rapidly drop off
as $b_N$ increases in a similar way. 
On the other hand, when the dinucleon is far from the core ($\beta = 5$ fm), 
the dinucleon probability amplitude of $^9$Li decreases a little more rapidly than that of $^9$C. 
This indicates that the dineutron and diproton keep almost the same compact size near the core, 
and that the diproton can swell more readily than the dineutron far from the core. 

In the following section, 
we discuss in detail the reasons why there is little quantitative difference for the dinucleon-size behavior near the core 
while differences can be evident far from the core. 
To do so, we investigate how large sized spin-singlet pairs of two nucleons are energetically favored 
under conditions of increasing spatial expansion from the core. 
We consider the dinucleon energy 
depending on the dinucleon size, $b_N$, and expansion, $\beta$, (where, by definition, $b_n < \beta$) 
in ideal $^9$Li ($\alpha+t+$dineutron) and $^9$C ($\alpha+h+$diproton) systems
by analyzing the energy of the corresponding DC wave functions. 
We again evaluate the dinucleon features by changing these parameters in the DC wave function. 
When $b_N$ is much smaller than $\beta$, a compact dinucleon is formed about the core. 
On the other hand, when $b_N \approx \beta$, 
each of two valence nucleons is in the single-particle $s$ wave about the core
and the dinucleon disperses because they do not have a spatial (angular) correlation. 
When $\beta$ is small, the two valence nucleons are distributed near the core, but,
when $\beta$ is large, they can be expanded widely. 
By calculating the energy change depending on the parameters $b_N$ and $\beta$, 
we see how a large dinucleon size is favored. 

The aim here is to discuss quantitative changes in the dinucleon size.  
For this aim, the $N$-$N$ interaction in the spin-singlet channel is essential. 
Therefore, in the following calculation, we use the values of $b=h=0.125$ 
which reproduce the unbound features 
of a spin-singlet, two-nucleon pair, in conjunction with $m=0.58$ and $v_{\rm LS} = 1600$ MeV. 

\begin{figure}[t]
\includegraphics[scale=0.7]{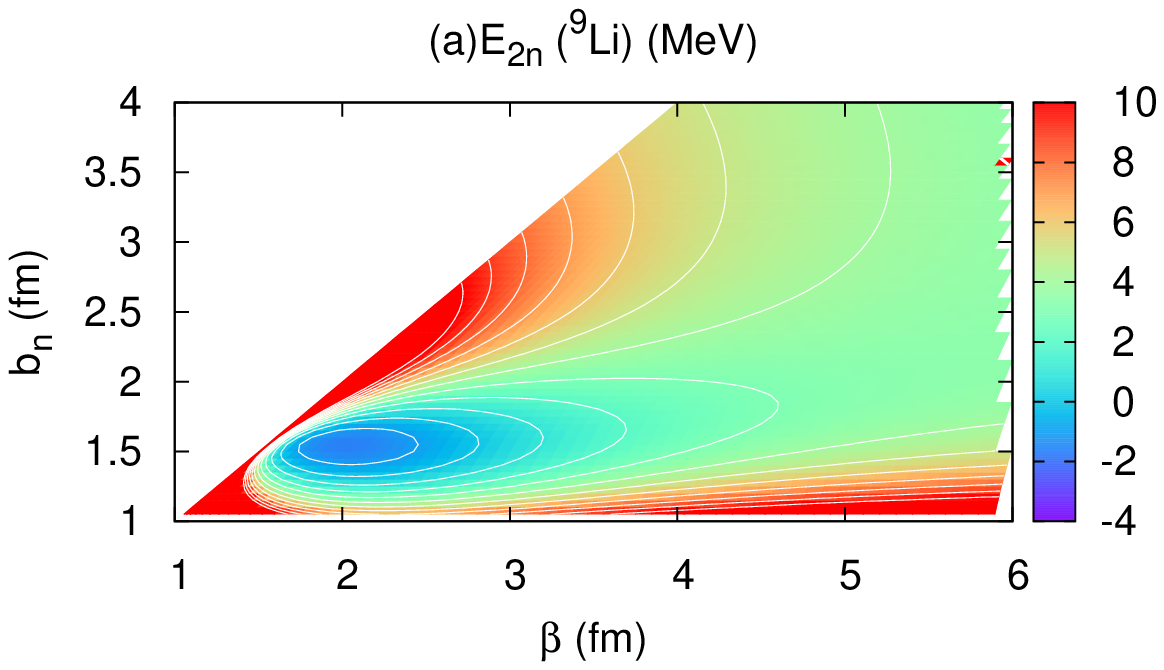} \\
\includegraphics[scale=0.7]{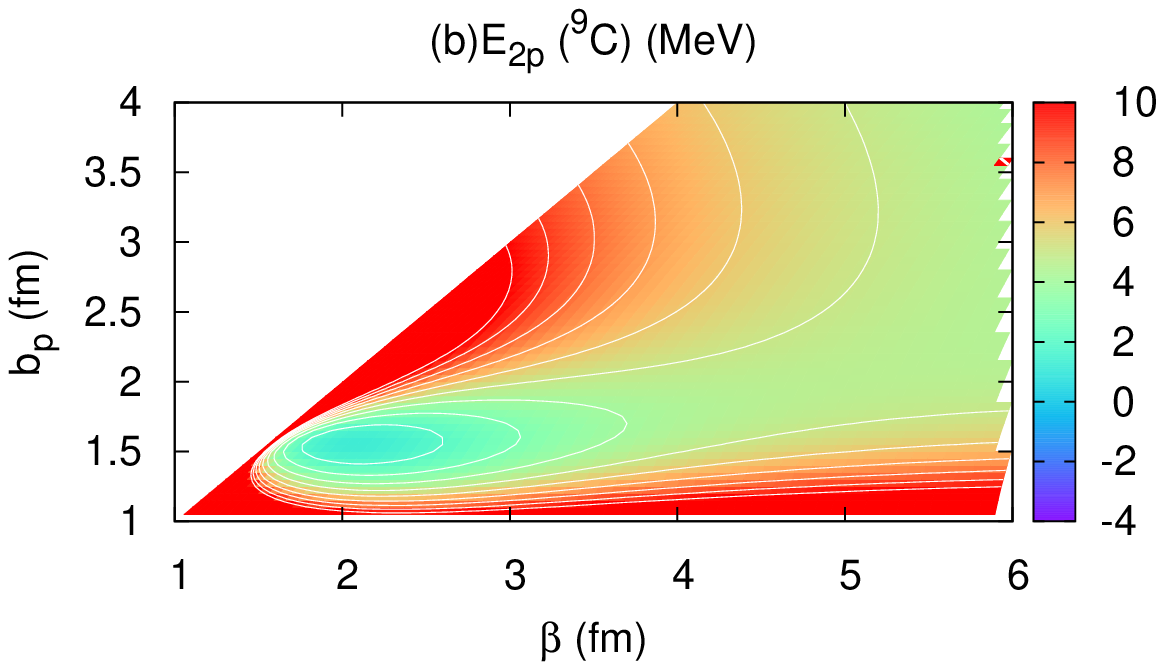}
\caption{(Color online) The $2N$-energy surface of the ideal $^9$Li and $^9$C systems on the $\beta$-$b_N$ plane. }
\label{fig:energy_surface_B_b}
\end{figure}

Here we calculate the energy of the $^9$C and $^9$Li DC wave functions 
and evaluate the $2N \ (N=p,n)$ energy 
with respect to the threshold energy of the core plus two free nucleons. 
\begin{align}
E_{2N}(\beta,b_N) = &\ \langle \Psi_{\rm DC}(\beta, b_N) |H| \Psi_{\rm DC}(\beta, b_N) \rangle \nonumber \\
&\ -  \langle \Psi_{\rm DC}^{\infty} |H| 
\Psi_{\rm DC}^{\infty} \rangle, 
\end{align}
where $|\Psi_{\rm DC} \rangle$ is the DC wave function for $^9$Li or $^9$C 
and $|\Psi_{\rm DC}^{\infty} \rangle$ is the DC wave function in the infinitely large $b_n$ and $\beta$ limit 
corresponding to the system of the core plus two free nucleons.  
We plot the energy surfaces of the ideal $^9$Li and $^9$C systems on the $\beta$-$b_N$ plane
in Fig.~\ref{fig:energy_surface_B_b}. 
Their absolute values are quite different over the entire region due to the Coulomb interaction, 
but the structures of their energy surfaces show similar features wherein 
an energy valley exists along the $b_N \sim 1.5$ fm line (corresponding to a compact dineutron)
and an energy barrier is formed in the region of $b_N \sim \beta$ 
(corresponding to two nucleons with little spatial correlation). 
Such an energy structure is the very dineutron downsizing mechanism 
discussed in our previous work \cite{kobayashi11}. 
The energy barrier is formed due to the Pauli repulsive effect from the core, 
and the energy valley is formed due to the attraction from the core. 
As a result of such an energy structure due to the core effect, 
a state containing a compact dinucleon is favored energetically, especially near the core. 
An energy barrier and valley exist for both $^9$Li and $^9$C; 
therefore, a compact dinucleon can be formed for both systems. 
We should note that, for realistic nuclei, 
the dinucleon dissociation due to the spin-orbit interaction also becomes significant 
depending on the core structure, as shown in Sec.~\ref{sec:dineutron}, 
and that, actually, formation and dissociation compete with each other. 
However, a compact dinucleon surely can be formed by this downsizing mechanism, 
at least for these light-mass nuclei. 

\begin{figure}
\begin{center}
\includegraphics[scale=0.6]{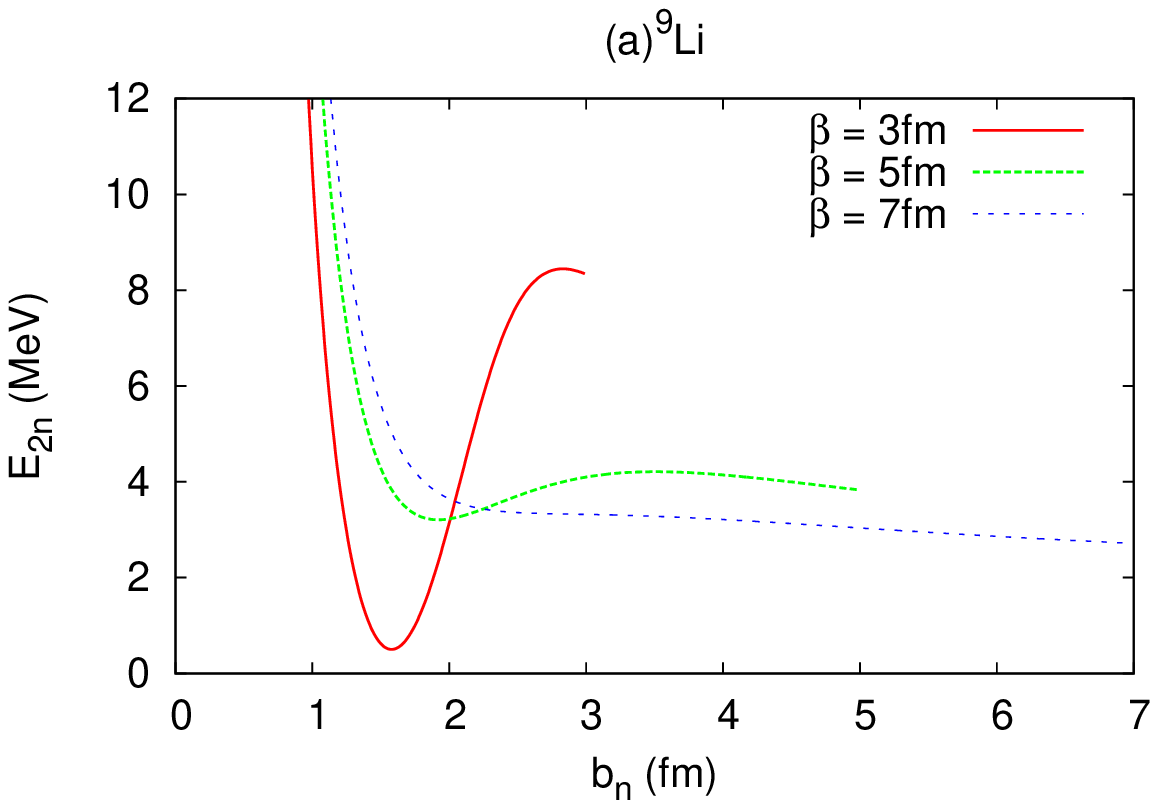} \\
\includegraphics[scale=0.6]{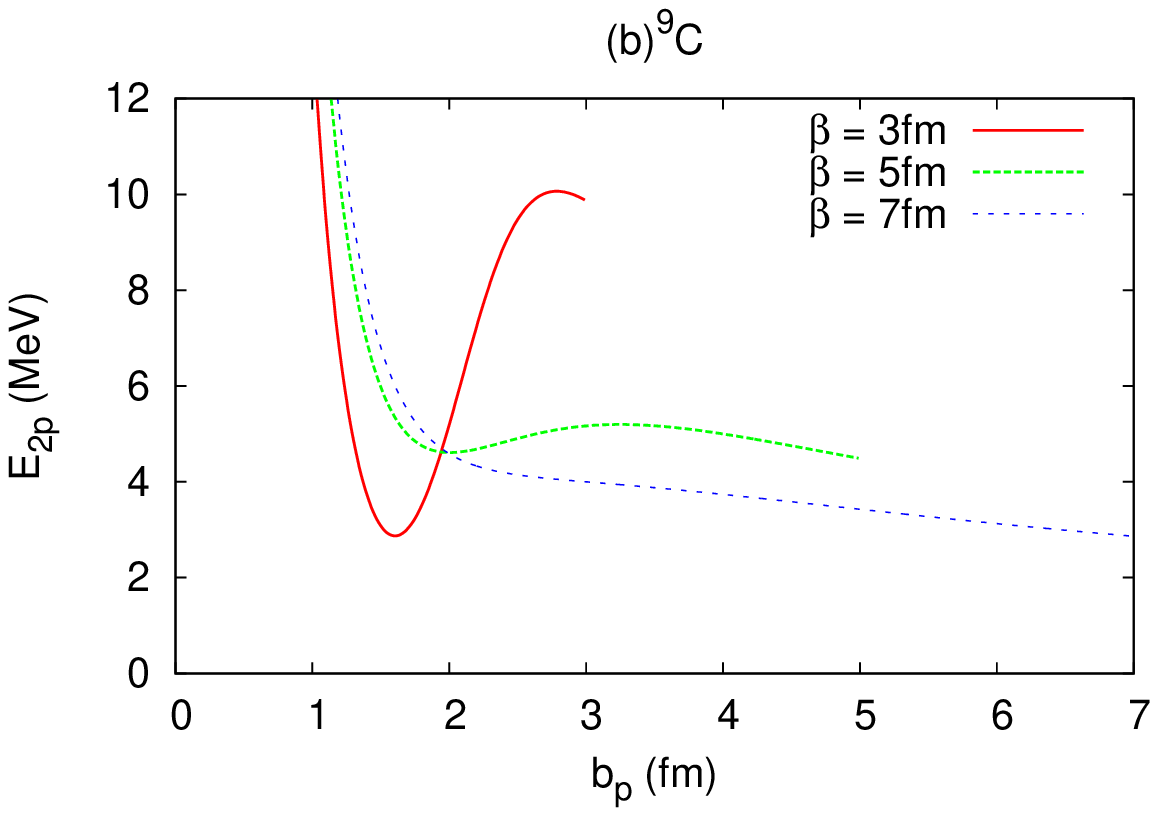}
\caption{(Color online) The $2N$ energy of the ideal $^9$Li and $^9$C systems as a function of $b_N$ 
where $\beta$ is fixed at $3$ (solid line), $5$ (dashed line), and $7$ fm (dotted line). }
\label{fig:energy_surface_b}
\end{center}
\end{figure}

To show the dinucleon-size dependence of the $2N$ energy more clearly, 
we plot the $2N$ energy as a function of $b_N$ with $\beta$ fixed to a certain value, which confines the valence nucleons to a finite region whose expansion is characterized by $\beta$. 
We therefore plot the $2N$ energy as a function of $b_N$ 
for $\beta = 3, 5$, and $7$ fm, as shown in Fig.~\ref{fig:energy_surface_b}. 
It can be clearly seen both for $^9$Li and $^9$C that an energy pocket exists at $b_N \sim 1.5$ fm 
and that a high energy barrier exists at $b_N \sim \beta$ 
when two valence neutrons are distributed near the core ($\beta = 3$ fm). 
The pocket becomes shallower and the barrier becomes lower in the $\beta = 5$ case 
and they disappear in the $\beta = 7$ case. 
This is because two valence nucleons can be distributed far from the core in the larger $\beta$ case 
so that the core effect, which is the basis of the observed energy pocket and barrier, becomes minor. 
The qualitative energy behavior is similar 
and, therefore, for $^9$C as well as $^9$Li, 
a compact dinucleon can be formed near the core despite the Coulomb force acting between the two protons. 

Although an energy barrier and pocket similarly form for both $^9$Li and $^9$C, 
the energy difference between the barrier (the local maximum at $b_N \sim \beta$)
and the pocket (the minimum at $b_N \sim 1.5-2$ fm) is somewhat different. 
The absolute values of the pocket and barrier for $^9$C 
are naturally higher than the corresponding values for $^9$Li due to the Coulomb repulsion. 
However, the energy difference is reduced for $^9$C by less than 1 MeV
compared with that for $^9$Li due to the Coulomb effect. 
When the dinucleon is distributed near the core (small $\beta$), 
the absolute value of the energy difference is large 
so that a reduction by only a few hundred keV has little contribution. 
On the other hand, when the dinucleon is expanded widely from the core (large $\beta$), 
the energy difference itself becomes lower 
and even a reduction by a few hundred keV can represent a significant contribution. 
For instance, in the case of $\beta = 5$ fm, 
the energy difference between the energy barrier and the pocket is $1.0$ MeV for $^9$Li 
and the difference reduces to about a half of this value for $^9$C. 
This means that, once the valence nucleons spread to regions further from the core, 
the difference between the diproton and dineutron correlation---that is, the tendency for a diproton to swell and disperse more readily than a dineutron---might be seen. 

 $^9$Li and $^9$C are not extremely loosely bound nuclei, 
in which two valence nucleons are distributed near the core. 
As a result, the core effect is influential to form strong dinucleon correlation, 
and the Coulomb effect has a minor contribution on the dinucleon correlation of the ground states of these nuclei. 
This is the reason why the differences between dineutron and diproton features are very small 
for $^9$Li and $^9$C, and, also, for $^{10}$Be and $^{10}$C, as discussed in Sec.~\ref{sec:dinucleon}. 
However, the size change can certainly become more significant for a diproton than a dineutron 
at further expansion from the core. 
The remarkable size change in a diproton may be seen distinctively
in extremely loosely bound proton-rich nuclei.

\section{Summary}
\label{sec:summary}

In this work, we have investigated dinucleon (dineutron and diproton) correlation 
about a core
in the ground states of $^9$Li, $^{10}$Be, and $^{9,10}$C 
using the cluster wave function and the DC wave function 
where core cluster breaking is taken into account. 

First, we have focused on $^9$Li and investigated the effect of $\alpha+t$ core structure change 
on dineutron correlation. 
In the situation that the $t$ cluster in the core can be broken, 
the neutrons in the core predominantly occupy the $0p_{3/2}$ orbits to gain the spin-orbit energy. 
The partial blocking of the $0p_{3/2}$ orbits by the core neutrons 
prevents the dineutron in the inner region from being dissociated into the $(0p_{3/2})^2$ configuration. 
As a result, two valence neutrons favor to form a compact spin-singlet dineutron inside the nuclei, 
and the dineutron probability in the inner region is markedly enhanced 
compared with the situation when the $t$ cluster cannot be broken. 
We have concluded that the occupation probability of the lower of the $ls$-splitting orbits in the core 
affect the degree of dineutron formation. 
In addition, we have explicitly shown that the spin-orbit interaction is essential 
for dineutron formation and dissociation, 
since a dineutron is fragile and dissociates readily due to the one-body spin-orbit interaction from the core. 
Next, we have investigated the effect of the $\alpha+\alpha$ core structure 
on the dineutron correlation of $^{10}$Be 
and compared the dineutron feature of $^{10}$Be with that of $^9$Li. 
An $\alpha$ cluster is more rigid and harder to be broken than a $t$ cluster. 
Therefore the dineutron enhancement mechanism mentioned above is ineffective for $^{10}$Be, 
and the dineutron probability does not increase for $^{10}$Be as it does for $^9$Li. 
Comparing $^9$Li and $^{10}$Be, 
the dineutron probability of $^{10}$Be is smaller over the entire region 
since the interaction between the core and valence neutrons for $^{10}$Be is stronger than that of $^9$Li 
and the dineutron is more attracted and dissociated in $^{10}$Be. 

We have also compared the diproton features of proton-rich $^{9,10}$C 
and the dineutron features of neutron-rich $^9$Li and $^{10}$Be. 
There are no qualitative differences between them 
and the Coulomb effect on dinucleon correlation is minor for these nuclei. 
We have discussed the quantitative differences between 
the degree of dinucleon size change for $^9$Li and $^9$C. 
We have shown that a dinucleon becomes similarly compact for both $^9$Li and $^9$C, 
but that the diproton can swell more readily than the dineutron due to the Coulomb effect. 
However, since the nuclei considered in the present work are not extremely loosely bound nuclei
and the core affects the dinucleon correlation strongly, 
the Coulomb effect on the dinucleon, which is inferior to the core effect, is minor in these nuclei. 

We have suggested that the dinucleon features, that
mainly consist of the degree of formation and the spatial expansion from the core, 
depend on the core structure. 
The occupied orbits by the core nucleons especially affect dinucleon correlation significantly. 
In the future, we will investigate dineutron correlation in neutron-rich Be and C isotopes 
to clarify the effect of the occupation probability of the lower orbits 
and also the effect of core deformation on dineutron correlation.

\appendix*
\section{Details of the measurement of dineutron probability}
\label{appendix}

In this appendix, we explain the detailed calculation of the existence probability 
of a compact dineutron, $\mathcal{N}_{\rm dineutron}$, in $^9$Li. 
In this calculation, 
we fix the dineutron size to a certain value, 
and $\mathcal{N}_{\rm dineutron}$ is reduced to a function of the distance ($d_{2n}$) 
between the $\alpha+t$ core and the dineutron
and the distance ($d_t$) between the $\alpha$ and $t$ clusters in the core. 
To calculate the dineutron probability, 
we prepare the wave function, $\Phi_{\rm dineutron}^{\lambda_t}$, describing the $\alpha+t$ core 
with the $\alpha$-$t$ distance of $d_t$  
plus a compact dineutron spherically distributed around the $\alpha+t$ core at a distance of $d_{2n}$, 
as follows. 
\begin{equation}
\Phi_{\rm dineutron}^{\lambda_t} (d_{2n}, d_t)
= \sum_i w_i \Phi_{\alpha+t+2n} (d_{2n}, \lambda_t, d_t; \Omega_i). 
\label{eq:dineutron_wf}
\end{equation}
In $\Phi_{\alpha+t+2n}$, the $\alpha+t$ core structure is characterized by the parameters 
$\lambda_t$ (the degree of $t$ cluster breaking) 
and $d_t$ (the $\alpha$-$t$ distance). 
The center of mass of the core is located at the origin. 
The Gaussian width of the nucleons in the core is $\nu = 0.235$ fm$^{-2}$. 
We express the spherical wave functions of two valence neutrons coupled to a spin singlet 
with Gaussian wave packets having the fixed width, $\nu = 0.22$ fm$^{-2}$ ($b = 1.5$ fm), 
which are located at the same position. 
The position labeled as $\Omega_i$ is a point distant from the origin by $d_{2n}$. 
We choose the apexes of the octahedron ($i=1, \ldots, 6$) 
and the cube ($i=7, \ldots, 14$) as $\Omega_i$. 
We take the angular average in the dineutron distribution around the core
by summing these 14 arrangements multiplied by a factor $w_i$. 
$w_i = 1/2 \times 1/6 \ (i=1, \ldots, 6)$ and $1/2 \times 1/8 \ (i=7, \ldots, 14)$. 
By using such factors, the wave function $\Phi_{\rm dineutron}^{\lambda_t}$ 
approximately describes the system of the $\alpha+t$ core plus the compact dineutron 
whose distribution is predominantly the $S$-wave around the core. 
We have checked that the results do not change 
even if we add the number of the dineutron position. 

By calculating the overlap with the wave function $\Phi^{\lambda_t}_{\rm dineutron}$, 
we estimate the dineutron probability 
that the compact dineutron is a distance $d_{2n}$ 
from the core of the $\alpha$ and $t$ clusters separated by a distance $d_t$. 
In the case of $^9$Li calculated without $t$ cluster breaking, 
we calculate the overlap with the wave function having $\lambda_t = 0.0$ fm
(Eq.~(\ref{eq:overlap_dineutron}) in the text). 
\begin{equation}
\mathcal{N}_{\rm dineutron}(d_{2n}, d_t)
= |\langle \Phi_{\rm dineutron}^{\lambda_t=0.0} (d_{2n}, d_t) | \Psi^{3/2^-_1}_M \rangle|^2. 
\label{eq:overlap_dineutron_0}
\end{equation}
In the case of $^9$Li calculated with $t$ cluster breaking, 
we introduce $t$ cluster breaking also in the test wave function 
of Eq.~(\ref{eq:dineutron_wf}). 
For each set of $(d_{2n}, d_t)$, 
we prepare a set of wave functions of $\Phi_{\rm dineutron}^{\lambda_t (n)}$ 
having $\lambda_t (n) = 0.0, 0.4$, and $0.8$ fm for $n = 1, 2, 3$. 
We orthogonalize these wave functions by a unitary matrix $U$ as
\begin{align}
\tilde{\Phi}_{\rm dineutron}^m (d_{2n}, d_t) 
= &\ \sum_n U_{mn} \Phi_{\rm dineutron}^{\lambda_t (n)} 
(d_{2n}, d_t) \nonumber \\ 
&\ (m, n = 1, 2, 3). 
\label{eq:dineutron_wf_orthogonalized}
\end{align} 
$\tilde{\Phi}_{\rm dineutron}^m$ span the subspace characterized by $\lambda_t$s. 
Calculating the summation of the overlaps with the orthogonalized wave functions, 
$\tilde{\Phi}_{\rm dineutron}^m$, 
we estimate the dineutron probability around the $\alpha+t$ core with $t$ cluster breaking. 
\begin{equation}
\mathcal{N}_{\rm dineutron}(d_{2n}, d_t)
= \sum_m |\langle \tilde{\Phi}_{\rm dineutron}^m (d_{2n}, d_t) | \Psi^{3/2^-_1}_M \rangle|^2. 
\label{eq:overlap_dineutron_lambda}
\end{equation}
We change the parameters $(d_{2n},d_t)$ in these overlaps 
[Eqs.~(\ref{eq:overlap_dineutron_0}) and (\ref{eq:overlap_dineutron_lambda})]
to investigate the core-dineutron distance 
and the $\alpha$-$t$ distance in the core. 
Also, for $^{10}$Be and $^{9,10}$C, we perform similar calculations as was done for $^9$Li.

\begin{acknowledgments}
This work was supported by a Grant-in-Aid for Scientific Research 
from the Japan Society for the Promotion of Science (JSPS).
This work was also supported by
a Grant-in-Aid for the Global COE Program ``The Next Generation of Physics,
Spun from Universality and Emergence'' 
from the Ministry of Education, Culture, Sports, Science and Technology (MEXT) of Japan.
A part of the computational calculations of this work was performed using the
supercomputers at YITP.
The authors would like to thank Enago (www.enago.jp) for the English language review.
\end{acknowledgments}

\bibliography{reference}

\end{document}